\documentclass[%
 reprint,
 amsmath,amssymb,
 aps,
pra,
]{revtex4-1}

\usepackage{amsthm}
\usepackage{graphicx}
\usepackage{dcolumn}
\usepackage{bm}
\usepackage{multirow} 
\usepackage{epstopdf}
\usepackage[ruled,boxed]{algorithm2e}
\usepackage{subfigure}
\usepackage{float}
\usepackage{booktabs}

\theoremstyle{remark}

\newtheorem{theorem}{\textbf{Theorem}}

\begin{document}

\makeatletter
\newcommand{\rmnum}[1]{\romannumeral #1}
\newcommand{\Rmnum}[1]{\expandafter\@slowromancap\romannumeral #1@}
\makeatother

\preprint{APS/123-QED}

\title{Estimating Gibbs partition function with quantum Clifford sampling}

\author{Yusen Wu$^{1}$}
\email{yusen.wu@research.uwa.edu.au}
\author{Jingbo Wang$^{1}$}
\email{jingbo.wang@uwa.edu.au}

\affiliation{%
 $^{1}$Department of Physics, The University of Western Australia, Perth, WA 6009, Australia\\
}%

\date{\today}

\begin{abstract}
The partition function is an essential quantity in statistical mechanics, and its accurate computation is a key component of any statistical analysis of quantum system and phenomenon. 
However, for interacting many-body quantum systems, its calculation generally involves summing over an exponential number of terms and can thus quickly grow to be intractable. Accurately and efficiently estimating the partition function of its corresponding system Hamiltonian then becomes the key in solving quantum many-body problems. In this paper we develop a hybrid quantum-classical algorithm to estimate the partition function, utilising a novel Clifford sampling technique.
Note that previous works on quantum estimation of partition functions require $\mathcal{O}(1/\epsilon\sqrt{\Delta})$-depth quantum circuits~\cite{Arunachalam2020Gibbs, Ashley2015Gibbs}, where $\Delta$ is the minimum spectral gap of stochastic matrices and $\epsilon$ is the multiplicative error. 
Our algorithm  requires only a shallow $\mathcal{O}(1)$-depth quantum circuit, 
repeated $\mathcal{O}(1/\epsilon^2)$ times,
to provide a comparable $\epsilon$ approximation. 
Shallow-depth quantum circuits are considered vitally important for currently available NISQ (Noisy Intermediate-Scale Quantum) devices.

\end{abstract}

\pacs{Valid PACS appear here}
\maketitle


\section{\label{sec:level1}Introduction}

Quantum computing makes use of quantum mechanical phenomena, such as quantum superposition and quantum entanglement, to perform computing tasks on quantum systems, which is fundamentally different from the classical computing \cite{Nielsen2002Quantum}. The most exciting thing about quantum computing is its ability to achieve significant speed-up over classical computing for solving certain problems, such as simulating quantum systems \cite{Childs2012Hamiltonian,Low2017Optimal}, factoring large integers \cite{Shor1997Polynomial}, random walk on graph models \cite{Szegedywalk,Wang2016Walk, Sam2020Walk, Wang2020Walk2, Xue2020Walk}, and unstructured database searching \cite{Grover1996A}. Unfortunately, the implementation of the most proposed quantum algorithms usually requires a fully functional quantum computer incorporating error correction \cite{Sergey2018Science}, that is beyond current experimental capabilities. In addition, near term quantum devices have limited qubits and a certain level of noise exists on each single- and double- qubit gate, therefore the gate noise will be accumulated with the increasing of the quantum circuit depth. Then finding out a practical computational task that shows quantum advantages on near term devices is of significance.

The \emph{partition function} is defined to describe the statistical properties of a physical system at a fixed inverse temperature. Nevertheless, the problem of computing the partition function of a physical system generally belongs to the $\#P$-hard complexity class \cite{LS10, Arunachalam2020Gibbs}. For example, Markov Chain Monte Carlo (MCMC) method \cite{Neal1993MCMC, Good2016DeepLearning, MCMC2009counting, Dyer1991Gibbs, LS10} provides an approach to sampling from high dimensional probability distributions. This method can be used to approximate partition functions with $\mathcal{O}(\Delta^{-1})$ sampling complexity, where $\Delta$ represents the spectral gap of stochastic matrices. If a stochastic matrix had an extremely small $\Delta$, it is extremely time-consuming to provide an estimation via the MCMC method. 

There has been several attempts in finding quantum algorithms to estimate partition functions, which are much more efficient than existing classical algorithms. These works used the techniques of phase estimation \cite{ManHong2011QMS,QMS2012,Ashley2015Gibbs,  Arunachalam2020Gibbs},  Szegedy quantum walk  \cite{Szegedywalk, MeanValue1, MeanValue2} or linear combinations of unitary method \cite{ Chow2016GibbsState} to provide an approximation of the partition function. Given the $\epsilon$ multiplicative error, these methods involved a $\mathcal{O}(\epsilon^{-1}\Delta^{-1/2})$-depth quantum circuit, achieving a polynomial speed-up in comparison with best known classical algorithms.  These algorithms thus require a fully functional quantum computer incorporating error correction in the small-scale $(\epsilon,\Delta)$ cases.


In order to achieve a significant quantum advantage in the Noisy Intermediate-Scale Quantum (NISQ) era, we design a classical-quantum hybrid algorithm to approximate the partition function of an arbitrarily complex Hamiltonian via using the quantum Clifford sampling technique \cite{Huang2020ClassicalShadow, Gross2015Clifford}. To the best of our knowledge, this is the first quantum algorithm for estimating partition functions using the Clifford sampling technique. The proposed method only requires a $\mathcal{O}(1)$-depth quantum circuit with an $(n+1)$-qubit quantum device to provide a comparable $\epsilon$ approximation of an $n$-qubit partition function. This substantial reduction in the circuit complexity is achieved by increasing  sampling complexity.  In other words, the $\mathcal{O}(1)$-depth quantum circuit needs to repeat  $\mathcal{O}(n/\epsilon^2)$ times to yield the $\epsilon$ approximation. In the following sections, a rigorous analysis of the quantum circuit complexity will be carried out, and the power of the proposed algorithm will be demonstrated by the fact that all numerical results are  within the expected $\epsilon$ multiplicative error.

\section{Theoretical background}

\subsection{Partition function of an $n$-qubit system}
For an $n$-qubit Hamiltonian $\mathcal{H}=\sum_{i=1}^Lh_i$ \cite{MacArdle2019VarITE}, where $h_i=P_1^{(i)}\otimes...\otimes P_n^{(i)}$ and $P_j^{(i)}\in\{I,\sigma^x,\sigma^y,\sigma^z\}$, its Gibbs state is defined as
\begin{align}
    |\mu_{\beta}\rangle=\mathcal{Z}(\beta)^{-1/2}\sum_{\mathbf{x}\in\Omega}e^{-\beta \mathcal{H}(\mathbf{x})/2}|\mathbf{x}\rangle
\end{align}
over the sample space (eigenvector space) $\mathbf{x}\in\Omega$, where $\mathbf{x}$ denotes one of the eigenvectors of $\mathcal{H}$, the corresponding eigenvalue $\mathcal{H}(\mathbf{x})=\langle\mathbf{x}|\mathcal{H}|\mathbf{x}\rangle$, and the real value $\beta$ is the inverse temperature. The partition function $\mathcal{Z}(\beta)$ is defined over the whole sample space, that is
\begin{align}
\mathcal{Z}(\beta)=\sum_{\mathbf{x}\in\Omega}\exp(-\beta \mathcal{H}(\mathbf{x})),
\end{align}
which contains an exponential number of terms and is therefore in general intractable computationally. 

\subsection{Quantum Clifford Sampling}
In quantum computation, the basic operators are the Pauli operators $\{I,\sigma^x,\sigma^y,\sigma^z\}$ which provide a basis for the density operators of a single qubit as well as for the unitaries that can be applied to them. For an $n$-qubit case, one can construct the Pauli group according to
$$\mathbf{P}_n=\{e^{i\theta\pi/2}\sigma^{j_1}\otimes...\otimes\sigma^{j_n}|j_k\in\{I,x,y,z\}\}.$$
Then the Clifford group $\mathbf{Cl}(2^n)$ is defined as the group of unitaries that normalize the Pauli group:
$$\mathbf{Cl}(2^n)=\{U|U\mathbf{P}_nU^{\dagger}=\mathbf{P}_n\},$$
and the Clifford gates are then defined as elements in the Clifford group, and these Clifford gates compose the Clifford circuit \cite{Gross2016Clifford}.

Randomly sampling Clifford circuits can reproduce the first $3$ moments of the full Clifford group endowed with the Haar measure $\mathbf{d}\mu_{{\rm{Haar}}}(U)\triangleq\mathbf{d}\mu(U)$ which is the unique left- and right- invariant measure such that
$$\int_{\mathbf{Cl}(2^n)}\mathbf{d}\mu(U)f(U)=\int\mathbf{d}\mu(U)f(VU)=\int\mathbf{d}\mu(U)f(UV)$$
for any $f(U)$ and $V\in\mathbf{Cl}(2^n)$. Using this property, one can sample Clifford circuits $U\in\mathbf{Cl}(2^n)$ with the probability ${\rm{Pr}}(U)$, and the corresponding expectation ${\rm{E}}_{U\in\mathbf{Cl}(2^n)}[\left(U\rho U^{\dagger}\right)^{\otimes t}]$ can be expressed as
$$\sum\limits_{U\in\mathbf{Cl}(2^n)}{\rm{Pr}}(U)\left(U\rho U^{\dagger}\right)^{\otimes t}=\int_{\mathbf{Cl}(2^n)}\mathbf{d}\mu(U)\left(U\rho U^{\dagger}\right)^{\otimes t}$$
for any $n$-qubit density matrix $\rho$ and $t=1,2,3$. The right hand side of the above equation can be evaluated explicitly by representation theory \cite{Gross2016Clifford}, this thus yields a closed-form expression for sampling from a Clifford group.

To extract meaningful information from a unknown quantum state $\rho$, the Clifford sampling technique was proposed by Huang et al \cite{Huang2020ClassicalShadow}. The Clifford sampling is implemented by repeatedly performing a simple measurement procedure: apply a random unitary $U\in \mathbf{Cl}(2^n)$ to rotate the state $\rho$ and perform a $\sigma^z$-basis measurement. The number of repeating times of this procedure is defined as the Clifford sampling complexity. On receiving the $n$-bit measurement outcome $|b\rangle\in\{0,1\}^n$, according to the Gottesman-Knill theorem \cite{Gottesman1997}, we can efficiently store an classical description of $U^{\dagger}|b\rangle\langle b|U$ in classical memory. This classical description encodes meaningful information of the state $\rho$ from a particular angle, and it is thus instructive to view the average mapping from $\rho$ to its classical snapshot $U^{\dagger}|b\rangle\langle b|U$ as a quantum channel:
\begin{align}
\mathcal{M}(\rho)=\textmd{E}_{U\in\mathbf{Cl}(2^n)}\left(\textmd{E}_{b\in\{0,1\}^n}[U^{\dagger}|b\rangle\langle b|U]\right),
\end{align}
where the quantum channel $\mathcal{M}$ depends on the ensemble of unitary transformation, and the quantum channel $\mathcal{M}$ can be further expressed as
\begin{align}
\mathcal{M}(\rho)=\textmd{E}_{U}\sum\limits_{\widehat{b}\in\{0,1\}^{n}}\langle \widehat{b}|U\rho U^{\dagger}|\widehat{b}\rangle U^{\dagger}|\widehat{b}\rangle\langle\widehat{b}|U=\frac{\rho+\rm{Tr}(\rho)I}{(2^n+1)2^n}.
\end{align}
Therefore the inverse of quantum channel $\mathcal{M}^{-1}(\rho)=(2^n+1)\rho-I$, and a \emph{Clifford sample} of $\rho$ is defined as
$$\widehat{\rho}=\mathcal{M}^{-1}\left(U^{\dagger}|b\rangle\langle b|U\right).$$
Repeat this procedure $M$ times results in an array of Clifford samples of $\rho$:
\begin{eqnarray}
\begin{split}
S(\rho;M)=\{&\widehat{\rho}_1=\mathcal{M}^{-1}\left(U_1^{\dagger}|b_1\rangle\langle b_1|U_1\right),...,\\
&\widehat{\rho}_M=\mathcal{M}^{-1}\left(U_M^{\dagger}|b_M\rangle\langle b_M|U_M\right)\},
\end{split}
\end{eqnarray}
which is defined as the \emph{Clifford Samples Set} of the quantum state $\rho$.
\section{Outline of the proposed Quantum-Classical hybrid algorithm}
In this section, we outline the fundamental $3$ steps of the proposed quantum-classical hybrid algorithm for computing $\mathcal{Z}(\beta)$ of a Hamiltonian $\mathcal{H}$, and these $3$ steps are named as the \emph{Partition-Function Clifford-Sampling (PFCS) Algorithm}:

Step 1 (CSBS). We propose a Clifford-Sampling-Binary-Search  algorithm to construct a sequence of increasing inverse temperatures $0=\beta_0<\beta_1<...<\beta_l=\beta$ which is called the \emph{cooling schedule}, and these temperatures satisfy
\begin{align}
c_1\leq\frac{\mathcal{Z}(\beta_i)\mathcal{Z}(\beta_{i+1})}{\mathcal{Z}(\frac{\beta_i+\beta_{i+1}}{2})^2}\leq c_2
\end{align}
for all $i\in\{0,...,l-1\}$ and two suitably chosen constants $c_1,c_2$.

Step 2 (PVGS). We propose a Projected-Variational-Gibbs-Sampling algorithm to calculate the quantum Gibbs state of a Hamiltonian. For all $\beta_i$ in the \emph{cooling schedule},
\begin{align}
|\mu_{\beta_i}\rangle=\sum\limits_{\mathbf{x}}\frac{e^{-\beta_i \mathcal{H}(\mathbf{x})/2}}{\sqrt{\mathcal{Z}(\beta_i)}}|\mathbf{x}\rangle,
\end{align}
where $\mathbf{x}$ denotes one of the eigenvectors of $\mathcal{H}$ and $\mathcal{H}(\mathbf{x})$ denotes the corresponding eigenvalue.

Step 3 (MECS). We provide the Mean-Value-Clifford-Sampling method. For $i\in\{0,...,l-1\}$, define random variables $V_i=\exp(-d_{i,i+1}\mathcal{H})$, and $W_i=\exp(d_{i,i+1}\mathcal{H})$, where $d_{i,i+1}=(\beta_{i+1}-\beta_i)/2$. After that, compute the expectation values of $V_i, W_i$:
$$\textmd{E}_{\mathbf{x}\sim\mu_{\beta_i}}[V_i]=\langle\mu_{\beta_i}|\exp\left(-d_{i,i+1}\mathcal{H}\right)|\mu_{\beta_i}\rangle=\frac{\mathcal{Z}(\frac{\beta_i+\beta_{i+1}}{2})}{\mathcal{Z}(\beta_i)},$$
and
$$\textmd{E}_{\mathbf{x}\sim\mu_{\beta_{i+1}}}[W_i]=\langle\mu_{\beta_{i+1}}|\exp\left(d_{i,i+1}\mathcal{H}\right)|\mu_{\beta_{i+1}}\rangle=\frac{\mathcal{Z}(\frac{\beta_i+\beta_{i+1}}{2})}{\mathcal{Z}(\beta_{i+1})},$$
then the partition function can be estimated as
\begin{align}
\mathcal{Z}(\beta)=\mathcal{Z}(\beta_0)\prod\limits_{i=0}^{l-1}\frac{\textmd{E}[V_i]}{\textmd{E}[W_i]}.
\label{Eq8}
\end{align}

\section{Clifford-Sampling-Binary-Search (CSBS) sub-algorithm
\label{Section4}
}

Here, we first indicate how many samples are sufficient to estimate the expectation of a product random variable with relative error. We will apply this result to perform the calculation of $\mathcal{Z}(\beta)$ given by Eq. (\ref{Eq8}). We then explain the necessity for designing the CSBS sub-algorithm to selecting a \emph{cooling schedule}. After that, we propose how to construct the CSBS algorithm via using quantum Clifford samplings.

For a random variable $X$, we use
\begin{align}
\textmd{S}[X]=\frac{\textmd{E}[X^2]}{(\textmd{E}[X])^2}
\end{align}
to represent the relative variance of $X$. Typically, Chebyshev's bound implies that at least $\mathcal{O}(\textmd{S}[X]/\epsilon_1^2)$ samples are required to estimate $\textmd{E}[X]$ with $\epsilon_1$ error. Therefore, if the relative variance $\textmd{S}[X]$ is extremely large (such as $\textmd{S}[X]={\rm{poly}}(n)$), the estimator is no longer efficient. 
\begin{theorem}
 Let $B>0$ and failure probability $\delta_1\in(0,1)$. Assume that the independent random variables $X_1,...,X_l$ satisfy $\textmd{S}[X_i]\leq B$ for all $i\in[l]$. By taking $m=2Bl/(\delta_1\epsilon_1^2)$ samples from $X_i$ for every $i\in[l]$, we can obtain $\widehat{X}=\prod_i{\rm{E}}[X_i]$ that satisfies
\begin{align}
\rm{Pr}\left[(1-\epsilon_1)\prod_i{\rm{E}}[X_i]\leq\widehat{X}\leq(1+\epsilon_1)\prod_i{\rm{E}}[X_i]\right]\geq1-\delta_1.
\end{align}

\label{Theorem1}
\end{theorem}

The proof of theorem \ref{Theorem1} utilizes Chebyshev's inequality~\cite{Dyer1991Gibbs,Arunachalam2020Gibbs}.
In the \emph{PFCS-Algorithm}, random variables $X_i$ take values $V_i$ and $W_i$ for $i\in[l]$, and their relative variances
\begin{align}
\textmd{S}[V_i]=\textmd{S}[W_i]=\frac{\mathcal{Z}(\beta_i)\mathcal{Z}(\beta_{i+1})}{\mathcal{Z}(\frac{\beta_i+\beta_{i+1}}{2})^2}.
\end{align}
According to the theorem \ref{Theorem1}, to efficiently estimate $\mathcal{Z}(\beta)$, we need to select a group of \emph{cooling schedule} $\beta_0<\beta_1<...<\beta_l$, where $\beta_0=0$ and $\beta_l=\beta$, whose relative variances are bounded by
$$c_1\leq\frac{\mathcal{Z}(\beta_i)\mathcal{Z}(\beta_{i+1})}{\mathcal{Z}(\frac{\beta_i+\beta_{i+1}}{2})^2}\leq c_2,$$
where $c_1$ and $c_2$ are two constants that are independent to the scale of the system $n$. Therefore, how to select a group of decent \emph{cooling schedule} is important for calculating the partition function, and the CSBS algorithm is thus proposed. The CSBS algorithm can be outlined as \emph{Algorithm1}, and details refer to the following two subsections.
\begin{algorithm}
\caption{CSBS Algorithm}\label{algorithm}
\textbf{Input:} Initial temperature $\beta_0=0$, largest temperature $\beta_l=\beta$, failure probability $\delta$, constant $c_2$.\\
\textbf{Output:} Set of cooling schedule $\beta_0,...,\beta_l$.\\
Set $k\leftarrow0$;\\
\textbf{while} $\beta_k<\beta_l$ \textbf{do}\\
(1) Invoking the \emph{Overlap Estimation} algorithm to compute the function $$f(\beta)=|\langle\mu_{\beta_k}|\mu_{\beta}\rangle|^2=\frac{\mathcal{Z}(\frac{\beta_k+\beta}{2})^2}{\mathcal{Z}(\beta_k)\mathcal{Z}(\beta)};$$\\
(2) Compute $\beta^*\leftarrow BinarySearch\left(f(\cdot)\geq c_2^{-1},[\beta_k,\beta_l],1/2n\right)$;\\
\textbf{return} $\beta_1,...,\beta_k$.
\label{Algorithm1}
\end{algorithm}
\subsection{Overlap Estimation}
According to the construction of estimators $V_i$ and $W_i$, we find that the inverse of their relative variances $S[V_i]^{-1}, S[W_i]^{-1}$ can be recognized as the quantum states overlap between Gibbs states $|\mu_{\beta_i}\rangle$ and $|\mu_{\beta_{i+1}}\rangle$, that is
\begin{align}
S[V_i]^{-1}=S[W_i]^{-1}=\frac{\mathcal{Z}(\frac{\beta_i+\beta_{i+1}}{2})^2}{\mathcal{Z}(\beta_i)\mathcal{Z}(\beta_{i+1})}=\left|\langle\mu_{\beta_i}|\mu_{\beta_{i+1}}\rangle\right|^2.
\end{align}
Therefore, one of the ingredients in CSBS algorithm relies on how to efficiently estimate the quantum states overlap. The \emph{Overlap Estimation Algorithm} is proposed as \emph{Algorithm2}, and the corresponding Clifford sampling complexity $M_s$ can be rigorously guaranteed by theorem \ref{Theorem2}.

\begin{algorithm}
\caption{Overlap Estimation Algorithm by sampling from $\textbf{Cl}(2^n)$ group}\label{algorithm}
\textbf{Input:} Quantum states $|\phi\rangle$, $|\psi\rangle$, accuracy parameter $\epsilon_2$, failure probability $\delta_2\in(0,1)$ and sampling complexity $M_s=\mathcal{O}(\log(1/\delta_2)\epsilon_2^{-2})$;\\
\textbf{Output:} Estimation of $|\langle\psi|\phi\rangle|^2$.\\
(1) Sampling $U$ from $\textbf{Cl}(2^n)$ group for $M_s$ times and construct the \emph{Clifford Samples Set} of the state $|\psi\rangle\langle\psi|$:
$$S(|\psi\rangle,M_s)=\{\widehat{\rho}_1(\psi),...,\widehat{\rho}_{M_s}(\psi)\}.$$ \\
(2) Sampling $U$ from $\textbf{Cl}(2^n)$ group for $M_s$ times and construct the \emph{Clifford Samples Set} of the state $|\phi\rangle\langle\phi|$:
$$S(|\phi\rangle,M_s)=\{\widehat{\rho}_1(\phi),...,\widehat{\rho}_{M_s}(\phi)\}.$$ \\
\textbf{return} 
$|\langle\psi|\phi\rangle|^2=\frac{1}{M_s}\sum\limits_{j=1}^{M_s}{\rm{Tr}}\left(\widehat{\rho}_j(\psi)\widehat{\rho}_j(\phi)\right)+\epsilon_2$.
\label{Algorithm2}
\end{algorithm}

\begin{theorem}
 Given two $n$-qubit quantum states $|\psi\rangle$, $|\phi\rangle$ and accuracy parameters $\epsilon_2, \delta_2\in[0,1]$, then a collection of $M_s=c\log(1/\delta_2)/\epsilon^2_2$ independent Clifford samplings suffice to estimate the overlap $|\langle\psi|\phi\rangle|^2$ with an additive error $\epsilon_2$ by using Alg. \ref{Algorithm2}, where $c$ is a constant value that is independent to $n$.
 \label{Theorem2}
\end{theorem}

\emph{Proof.} Using $\widehat{o}(M_s,\psi,\phi)$ to represent the estimation value $\sum_{j=1}^{M_s}{\rm{Tr}}\left(\widehat{\rho}_j(\psi)\widehat{\rho}_j(\phi)\right)/M_s$ and $o(\psi,\phi)$ to represent the exact value of $|\langle\psi|\phi\rangle|^2$, then according to Hoeffding's inequality, the failure probability $\delta$ can be estimated as
$${\rm{Pr}}\left(|\widehat{o}(M_s,\psi,\phi)-o(\psi,\phi)|\geq\epsilon_2\right)\leq\exp\left(\frac{-2\ln2M_s\epsilon_2^2}{{\rm{Var}}(\widehat{o}(M_s,\psi,\phi))}\right),$$
where ${\rm{Var}}(\widehat{o}(M_s,\psi,\phi))$ represents the variance of the estimation algorithm, therefore
\begin{align}
M_s=\mathcal{O}\left(\frac{{\rm{Var}}(\widehat{o}(M_s,\psi,\phi))\log(1/\delta_2)}{\epsilon_2^2}\right).
\end{align}
According to the Lemma 1 in the literature \cite{Huang2020ClassicalShadow}, the variance ${\rm{Var}}(\widehat{o}(M_s,\psi,\phi))$ can be estimated by
\begin{eqnarray}
\begin{split}
&\max_{\sigma}\textmd{E}_{U\sim Cl(2^n)}\sum\limits_{b\in\{0,1\}^n}\langle b|U\sigma U^{\dagger}|b\rangle\langle b|U\mathcal{M}^{-1}(O_{\psi})U^{\dagger}|b\rangle^2\\
&=\max_{\sigma}{\rm{Tr}}\left(\sigma\sum\limits_{b\in\{0,1\}^n}\frac{(2^n+1)^2({\rm{Tr}}(O_{\psi}^2)I+2O_{\psi}^2)}{(2^n+2)(2^n+1)2^n}\right)\\
&=\frac{2^n+1}{2^n+2}\max_{\sigma}\left({\rm{Tr}}(\sigma){\rm{Tr}}(O_{\psi}^2)+2{\rm{Tr}}(\sigma O_{\psi}^2)\right)\leq c,
\end{split}
\end{eqnarray}
where $O_{\psi}=|\psi\rangle\langle\psi|-I/2^n$ and $c$ is a constant value that is independent to the scale of the quantum system $n$. Combining the above two equations, we can obtain the lower bound of quantum sampling complexity $M_s=c\log(1/\delta_2)/\epsilon^2_2$. $\Box$

Actually, theorem \ref{Theorem2} indicates that one can efficiently estimate the function $f(\beta)=|\langle\mu_{\beta_k}|\mu_{\beta}\rangle|^2$ which directly reflects the variance of the cooling schedule without using any ancillary qubit. Compared with the previous arts \cite{Ashley2015Gibbs, Arunachalam2020Gibbs} that invoke the amplitude estimation algorithm, the Alg. \ref{Algorithm2} does not need a $\mathcal{O}(1/\epsilon_2+n^2)$ depth quantum circuit, but a $(n^2/\log n)$-depth random Clifford quantum circuit suffices to estimate the value of $|\langle\psi|\phi\rangle|^2$. In addition, one can further modify the Alg. \ref{Algorithm2} by only sampling from $\mathbf{Cl}(2^k)$ group ($k<n$). According to ``no free lunch"  theorem, this modification must introduces additional quantum sampling complexity. The corresponding algorithm is shown as Alg. \ref{Algorithm3}, and the corresponding quantum sampling complexity $M_s$ can be rigorously guaranteed by the following Theorem.

\begin{algorithm}
\caption{Overlap Estimation Algorithm by sampling from $\mathbf{Cl}(2^k)$ group}\label{algorithm}
\textbf{Input:} Quantum states $|\phi\rangle$, $|\psi\rangle$, accuracy parameters $\epsilon_2$, failure probability $\delta_2\in[0,1]$ and sampling complexity $M_s$;\\
\textbf{Output:} Estimation of $|\langle\psi|\phi\rangle|^2$.\\
(1) Suppose $U=\bigotimes_{j=1}^{[n/k]}U_j$ and each $U_j$ is sampled from $\mathbf{Cl}(2^k)$ group. Repeat this procedure for $M_s$ times and construct Classical Shadow sets of the state $|\psi\rangle\langle\psi|$:
$$S(|\psi\rangle,M_s)=\{\widehat{\rho}_1(\psi),...,\widehat{\rho}_{M_s}(\psi)\},$$ \\
where
$$\widehat{\rho}_i(\psi)=\mathcal{M}^{-1}(U^{\dagger}|\widehat{b}\rangle\langle\widehat{b}|U)=\bigotimes\limits_{j=1}^{[n/k]}\left((2^k+1)U_j^{\dagger}|\widehat{b}_j\rangle\langle\widehat{b}_j|U_j-I\right).$$\\
(2) Suppose $U=\bigotimes_{j=1}^{[n/k]}U_j$ and each $U_j$ is sampled from $\mathbf{Cl}(2^k)$ group. Repeat this procedure for $M_s$ times and construct Classical Shadow sets of the state $|\phi\rangle\langle\phi|$:
$$S(|\phi\rangle,M_s)=\{\widehat{\rho}_1(\phi),...,\widehat{\rho}_{M_s}(\phi)\},$$ \\
\textbf{return}
$|\langle\psi|\phi\rangle|^2=\frac{1}{M_s}\sum\limits_{j=1}^{M_s}{\rm{Tr}}\left(\widehat{\rho}_j(\psi)\widehat{\rho}_j(\phi)\right)+\epsilon_2$.
\label{Algorithm3}
\end{algorithm}

\begin{theorem}
Given two $n$-qubit quantum states $|\psi\rangle$, $|\phi\rangle$ and accuracy parameters $\epsilon_2, \delta_2\in(0,1)$, then a collection of 
\begin{align}
M_s=\left(\frac{3(2^k+1)}{(2^k+2)}\right)^{[n/k]}\frac{\log(1/\delta_2)}{\epsilon^2_2}
\end{align}
independent Clifford samplings suffice to estimate the overlap $|\langle\psi|\phi\rangle|^2$ with an additive error $\epsilon_2$ by $\mathbf{Cl}(2^k)$ sampling.
\label{Theorem4}
\end{theorem}

\emph{Proof.} Still using $\widehat{o}(M_s,\psi,\phi)$ to represent the estimation value $\sum_{j=1}^{M_s}{\rm{Tr}}\left(\rho_j(\psi)\rho_j(\phi)\right)/M_s$, the variance ${\rm{Var}}(\widehat{o}(M_s,\psi,\phi))$ can be estimated as
\begin{widetext}
\begin{equation}
\begin{split}
    &\max_{\sigma}\textmd{E}_{U\sim \mathbf{Cl}(2^k)^{\otimes [n/k]}}\sum\limits_{b\in\{0,1\}^n}\langle b|U\sigma U^{\dagger}|b\rangle\langle b|U\mathcal{M}^{-1}(O_{\psi})U^{\dagger}|b\rangle^2\\
    &=\max_{\sigma}{\rm{Tr}}\left(\sigma\bigotimes\limits_{j=1}^{n/k}\textmd{E}_{U_j\sim \mathbf{Cl}(2^k)}\sum\limits_{b_j=0}^{2^k-1}U^{\dagger}_j|b_j\rangle\langle b_j|U_j\langle b_j|U_jO_{\psi}U_j^{\dagger}|b_j\rangle^2\right)\\
    &=\max_{\sigma}{\rm{Tr}}\left(\sigma\bigotimes\limits_{j=1}^{n/k}\sum\limits_{b_j=0}^{2^k-1}\frac{(2^k+1)^2\left({\rm Tr}(O_{\psi_j}^2)I+2O_{\psi_j}^2\right)}{(2^k+2)(2^k+1)2^k}\right)\\
    &\leq\left(\frac{3(2^k+1)}{2^k+2}\right)^{[n/k]},
    \end{split}
\end{equation}
\end{widetext}
in which $O_{\psi}=|\psi\rangle\langle\psi|-I/2^n$, $O_{\psi_j}=3(|\psi_j\rangle\langle\psi_j|-I/2^k)$, and $|\psi_j\rangle$ indicates the qubits performed by $U_j$. Combing the Hoeffding's inequality, one can obtain the required sampling complexity in Alg. \ref{Algorithm3}. $\Box$

If we choose $k=1$, the Alg. \ref{Algorithm3} degenerates to the single-qubit sampling algorithm, and the sampling complexity is shown as theorem \ref{Theorem4}.

\begin{theorem}
 Given two $n$-qubit quantum states $|\psi\rangle$, $|\phi\rangle$ and accuracy parameters $\epsilon_2, \delta_2\in(0,1)$, then a collection of $M_s=2.25^n\log(1/\delta_2)/\epsilon^2_2$ independent Clifford samplings suffice to estimate the overlap $|\langle\psi|\phi\rangle|^2$ with an additive error $\epsilon_2$ by sampling from $\mathbf{Cl}(2)$ group.
\label{Theorem4}
\end{theorem}

\subsection{Binary Search Algorithm}
The $BinarySearch$ algorithm aims at finding a subinterval $[\beta_k,\beta_{k+1}]$ from the large interval $[\beta_k,\beta]$ ($\beta_{k+1}\leq\beta$) that enables the relative variance
$$\frac{1}{f(\beta_{k+1})}=\frac{\mathcal{Z}(\beta_k)\mathcal{Z}(\beta_{k+1})}{\mathcal{Z}(\frac{\beta_k+\beta_{k+1}}{2})^2}\leq c_2,$$
where $c_2$ is a constant value. To do this, we introduce a monotone predicate $\mathcal{P}(f(\beta))$. A monotone predicate $\mathcal{P}$ is a boolean function defined on a totally ordered set with the property: if $\mathcal{P}(f(x))=\emph{true}$, then $\mathcal{P}(y)=\emph{ture}$ for all $y\leq x$ in the domain. In our case, $\mathcal{P}(f(\beta))$ returns \emph{true} at $\beta$
but returns \emph{false} at $\beta+1/{\rm{poly}}(n)$ when relationships
$f(\beta)\geq c^{-1}_2$ and $f(\beta+1/{\rm{poly}}(n))<c_2^{-1}$ hold at the same time, and the $BinarySearch$ algorithm is illustrated as follows.

\begin{algorithm}
\caption{Binary Search Algorithm \cite{Arunachalam2020Gibbs}}\label{algorithm}
\textbf{Input:} Monotone predicate $\mathcal{P}$, interval $[\beta_k,\beta_l]$ such that $\mathcal{P}(\beta_k)=\emph{true}$, precision $\alpha$.\\
\textbf{Output:} $\beta_l$ if $\mathcal{P}(\beta_l)=true$, otherwise an $\beta$ such that $\mathcal{P}(\beta)=\emph{true}$ and $\mathcal{P}(\beta+\alpha)=\emph{false}$.\\

\textbf{if} $\mathcal{P}(\beta_l)=\emph{true}$ \textbf{then}

\qquad\textbf{return} $\beta_l$\\

Set $\beta\leftarrow \beta_k$, $s\leftarrow \beta_l$;\\
\textbf{while} $s-\beta>\alpha$ \textbf{do}

\qquad\textbf{if} $\mathcal{P}(\frac{s+\beta}{2})=\emph{true}$ \textbf{then}

\qquad\qquad$\beta\leftarrow \frac{s+\beta}{2}$

\qquad\textbf{else}

\qquad\qquad $s\leftarrow \frac{s+\beta}{2}$



\textbf{return} $\beta$
\label{Algorithm4}
\end{algorithm}

\section{Projected-Variational-Gibbs-Sampling (PVGS)
\label{Section5}}
 In this section, we propose a shallow-circuit algorithm to complete the second step in the \emph{PFCS-Algorithm}, that is, preparing a quantum Gibbs state
$$|\mu_{\beta}\rangle=\sum\limits_{\mathbf{x}}\frac{e^{-\beta \mathcal{H}(\mathbf{x})/2}}{\sqrt{\mathcal{Z}(\beta)}}|\mathbf{x}\rangle|\mathbf{x}\rangle
$$
for an inverse temperature $\beta$ and Hamiltonian $\mathcal{H}=\sum_{\mathbf{x}}\mathcal{H}(\mathbf{x})|\mathbf{x}\rangle\langle\mathbf{x}|$.
\begin{figure*}
  \begin{center}
  \includegraphics[width=0.80\textwidth]{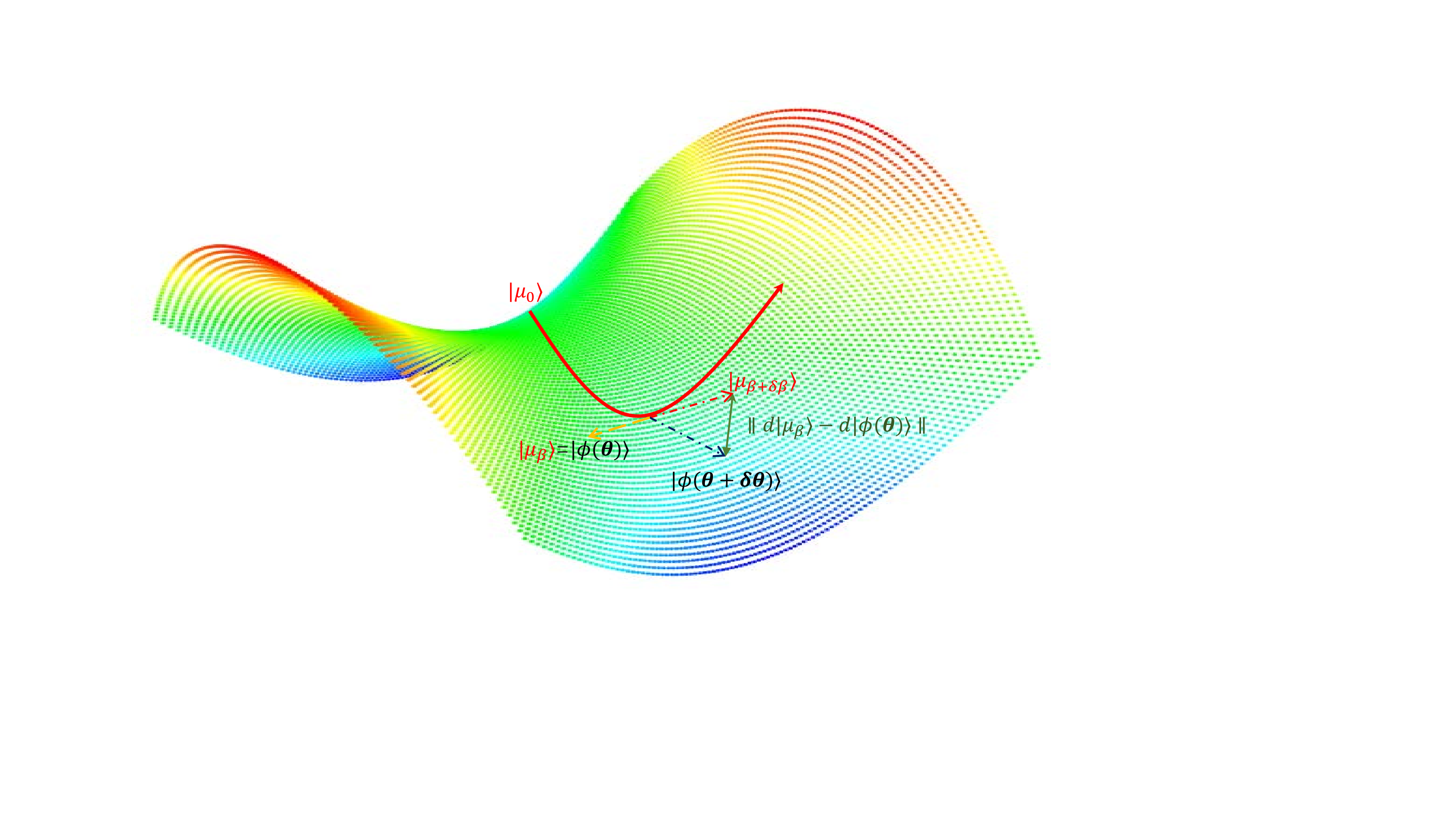}\\
  \caption{The schematic diagram of the imaginary time evolution manifold. The red solid arrow indicates the ideal path of $|\mu_{\beta}\rangle$, and the red dotted arrow indicates the variation of $|\mu_{\beta}\rangle$ at the point $\beta+\delta\beta$. The black dotted arrow represents the variation of $|\phi(\bm{\theta})\rangle$ at the point $(\bm{\theta}+\delta\bm{\theta})$, and the green double-headed arrow represents the difference two variations (see equation (\ref{Eq:difference})).}
  \end{center}
  \label{Fig:RBM}
\end{figure*}

To do this, we first prepare a initial state
\begin{align}
|\mu_0\rangle=\frac{1}{2^{n/2}}\sum_{\mathbf{i}}|\mathbf{i}\rangle|\mathbf{i}\rangle=\frac{1}{2^{n/2}}\sum_{\mathbf{x}}|\mathbf{x}\rangle|\mathbf{x}\rangle \end{align}
via performing $n$-qubit Hadamard gate $H^{\otimes n}$ and a series of CNOT gate onto the state $|0\rangle^{\otimes n}|0\rangle^{\otimes n}$. After that, we can perform $e^{-\beta\mathcal{H}/2}$ onto the initial state $|\mu_0\rangle$, that is
\begin{eqnarray}
\begin{split}
|\mu_{\beta}\rangle=\frac{\exp(-\frac{\beta}{2}\mathcal{H})|\mu_0\rangle}{\sqrt{\langle\mu_0|e^{-\beta\mathcal{H}}|\mu_0\rangle}}.
\end{split}
\end{eqnarray}
This procedure is also named as imaginary time evolution and the relevant practical quantum algorithms have been proposed in literatures \cite{MacArdle2019VarITE, Mario2019QITE}. These algorithms are based on a reformulation of the Dirac-Frenkel and McLachlan variational principle, called the Time-Dependent Variational Principle (TDVP). The TDVP-based algorithms iteratively update the variational parameters via Euler method, and this kind of algorithm thus losts high-order information of the variational parameters. To tackle this problem, we propose another method for implementing $|\mu_{\beta}\rangle$ via directly calculating the variation of parameters.
\subsection{Outline of the PVGS}
Instead of directly encoding the quantum state $|\mu_{\beta}\rangle$ at inverse temperature $\beta$, we approximate it using a parameterized trial state $|\phi(\bm{\theta})\rangle$, with $\bm{\theta}=(\theta_1,\theta_2,...,\theta_D)$. This stems from the intuition that the physically relevant state are contained in a small subspace of the full Hilbert space. The trial state is referred to as the ansatz. In condensed matter physics and computational chemistry, a wide variety of ansatz have been proposed for both classical and quantum variational methods. Using a quantum circuit, we prepare the trial state by
\begin{align}
|\phi(\bm{\theta})\rangle=\prod\limits_{d=1}^D\widetilde{U}_d(\theta_d)|\mu_0\rangle,
\end{align}
where $\widetilde{U}_d(\theta_d)=U_d(\theta_d)W_d$. The notation $U_d(\theta_d)$ is the unitary gate (single- or double-qubit gate), controlled by parameter $\theta_d$, and $W_d$ is the double-qubit gate independent to $\theta_d$.

Suppose the state $|\mu_{\beta}\rangle$ at inverse temperature $\beta$ is approxiamted by the trial state $|\phi(\bm{\theta})\rangle$ with parameters $\bm{\theta}$, then we want to approximate the state $|\mu_{\beta+\delta\beta}\rangle$ at inverse temperature $\beta+\delta\beta$ by $|\phi(\bm{\theta}+\delta\bm{\theta})\rangle$. The value of $\delta\bm{\theta}=(\delta\bm{\theta}_1,\delta\bm{\theta}_2,...,\delta\bm{\theta}_D)$ can be determined by minimizing the distance
\begin{align}
\mathcal{L}(\delta\bm{\theta})=\|\mathbf{d}|\mu_{\beta+\delta\beta}\rangle-\mathbf{d}|\phi(\bm{\theta}+\delta\bm{\theta})\rangle\|,
\label{Eq:difference}
\end{align}
where
\begin{align}
\mathbf{d}|\mu_{\beta+\delta\beta}\rangle=\frac{e^{-\delta\beta\mathcal{H}}|\mu_{\beta}\rangle}{\sqrt{\langle\mu_{\beta}|e^{-2\delta\beta\mathcal{H}}|\mu_{\beta}\rangle}}-|\mu_{\beta}\rangle, \end{align}
\begin{align}
\mathbf{d}|\phi(\bm{\theta}+\delta\bm{\theta})\rangle=\sum\limits_{d=1}^D\frac{\partial|\phi(\bm{\theta})\rangle}{\partial\bm{\theta}_d}\delta\bm{\theta}_d,
\end{align}
and the notation $\|\cdot\|$ represents the fidelity norm. Then the function $\mathcal{L}(\delta\bm{\theta})$ can be further computed as
\begin{eqnarray}
\begin{split}
&\mathcal{L}^2(\delta\bm{\theta})=\mathbf{d}\langle\mu_{\beta+\delta\beta}|\mathbf{d}|\mu_{\beta+\delta\beta}\rangle-\sum\limits_{d=1}^D\mathbf{d}\langle\mu_{\beta+\delta\beta}|\frac{\partial|\phi(\bm{\theta})\rangle}{\partial\bm{\theta}_d}\delta\bm{\theta}_d\\
&-\sum\limits_{d=1}^D\frac{\partial\langle\phi(\bm{\theta})|}{\partial\bm{\theta}_d}\mathbf{d}|\mu_{\beta+\delta\beta}\rangle\delta\bm{\theta}_d+\sum\limits_{m,n}\frac{\partial\langle\phi(\bm{\theta})|}{\partial\bm{\theta}_m}\frac{\partial|\phi(\bm{\theta})\rangle}{\partial\bm{\theta}_n}\delta\bm{\theta}_m\delta\bm{\theta}_n.
\end{split}
\end{eqnarray}
If we focus on the $m$-th variable $\delta\bm{\theta}_m$, the minimum of $\mathcal{L}^2(\delta\bm{\theta})$ obtains at
\begin{align}
\sum\limits_{m=1}^DA_{n,m}\delta\bm{\theta}_m=C_m,
\end{align}
in which the parameter
\begin{align}
A_{n,m}=\Re\left(\frac{\partial\langle\phi(\bm{\theta})|}{\partial\bm{\theta}_n}\frac{\partial|\phi(\bm{\theta})\rangle}{\partial\bm{\theta}_m}\right),
\end{align}
and
\begin{align}
C_m=\Re\left(\frac{\partial\langle\phi(\bm{\theta})|}{\partial\bm{\theta}_m}\mathbf{d}|\mu_{\beta+\delta\beta}\rangle\right).
\end{align}
Once each elements are provided, the change of parameters $\delta\bm{\theta}$ can be efficiently computed by solving the linear system
\begin{align}
A(\bm{\theta})\delta\bm{\theta}=C(\bm{\theta}),
\end{align}
where the matrix $A(\bm{\theta})=(A_{n,m})_{D\times D}$ and $C(\bm{\theta})=(C_1,...,C_D)^T$. Since the matrix $A$ is a real-valued symmetry matrix, the inverse of $A$ must exist. And $\bm{\theta}$ can be updated by
\begin{align}
\bm{\theta}+\delta\bm{\theta}=\bm{\theta}+A^{-1}(\bm{\theta})C(\bm{\theta}).
\end{align}
Finally, the Gibbs state $|\mu_{\beta+\delta\beta}\rangle$ can be approximated by $|\phi(\bm{\theta}+\delta\bm{\theta})\rangle$.
\begin{figure*}
  \begin{center}
  \includegraphics[width=0.80\textwidth]{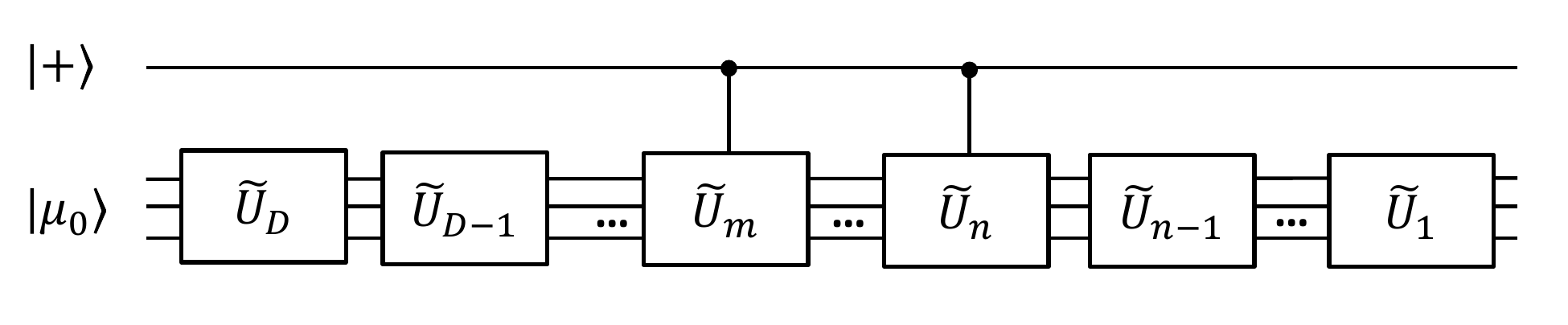}\\
  \caption{The quantum circuit for implementing controlled unitary (Eq.(\ref{Eq:ControlledRotation})).}
  \end{center}
  \label{Fig:ControlledRotation}
\end{figure*}
\subsection{Technical details for estimating $A(\bm{\theta})$ and $C(\bm{\theta})$}
Now we provide details on how to estimate each elements in matrix $A(\bm{\theta})$ and vector $C(\bm{\theta})$.

The element $A_{n,m}$ can be recognized as the real part of the inner-product between two quantum states $\frac{\partial\langle\phi(\bm{\theta})|}{\partial\bm{\theta}_n}$ and $\frac{\partial|\phi(\bm{\theta})\rangle}{\partial\bm{\theta}_m}$. The quantum state
\begin{align}
\frac{\partial|\phi(\bm{\theta})\rangle}{\partial\bm{\theta}_m}=\widetilde{U}_{D:m+1}U(\theta_m-\frac{\pi}{2})W_m\widetilde{U}_{m-1:1}|\mu_0\rangle,
\end{align}
in which the notation $\widetilde{U}_{j:i}=\prod_{s=j}^i\widetilde{U}_s(\theta_s)$. Apparently, the state $\frac{\partial|\phi(\bm{\theta})\rangle}{\partial\bm{\theta}_m}$ can be obtained by directly shifting the parameter $\theta_m \leftarrow\theta_m-\pi/2$. Then one can utilize the Hardamard Test algorithm to estimate the value of $A_{n,m}$. To do this, one need to perform the controlled unitary operator
\begin{eqnarray}
\begin{split}
|0\rangle\langle0|&\otimes\widetilde{U}_{D:m+1}U(\theta_m-\frac{\pi}{2})W_m\widetilde{U}_{m-1:1}\\
&+|1\rangle\langle1|\otimes\widetilde{U}_{D:n+1}U(\theta_n-\frac{\pi}{2})W_n\widetilde{U}_{n-1:1}
\end{split}
\label{Eq:ControlledRotation}
\end{eqnarray}
onto the state $|+\rangle|\mu_0\rangle$, the system thus becomes to
\begin{align}
\frac{1}{\sqrt{2}}\left(|0\rangle\frac{\partial|\phi(\bm{\theta})\rangle}{\partial\bm{\theta}_m}+|1\rangle\frac{\partial|\phi(\bm{\theta})\rangle}{\partial\bm{\theta}_n}\right).
\end{align}
Specifically, since the circuit structure of $\frac{\partial|\phi(\bm{\theta})\rangle}{\partial\bm{\theta}_m}$ is similar to $\frac{\partial|\phi(\bm{\theta})\rangle}{\partial\bm{\theta}_n}$, we can implement the controlled unitary (Eq.(\ref{Eq:ControlledRotation})) by only using two controlled unitaries (see Fig.(2)).
Then, we perform the Hardamard gate $H$ onto the first qubit, and measure the first qubit via Pauli Z basis, the value of $A_{n,m}$ can thus be estimated by
\begin{align}
\widehat{A}_{n,m}=2\Pr(0)-1,
\label{Eq:Hardamard-Test}
\end{align}
where $\Pr(0)$ is the probability for measuring the $|0\rangle$ state.

Similarly, the element $C_m$ can be recognized as the real part of the inner-product between quantum states $\frac{\partial|\phi(\bm{\theta})\rangle}{\partial\bm{\theta}_m}$ and $\mathbf{d}|\mu_{\beta+\delta\beta}\rangle$. Since $|\mu_{\beta}\rangle$ can be approximated by $|\phi(\bm{\theta})\rangle$ at inverse temperature $\beta$, then $C_m$ can be further expressed as
\begin{eqnarray}
\begin{split}
C_m= &\left(\frac{1}{\sqrt{E_\beta}}-1\right)\Re\left(\langle\phi(\bm{\theta})|\frac{\partial|\phi(\bm{\theta})\rangle}{\partial\bm{\theta}_m}\right)\\
-&\frac{\delta\beta}{\sqrt{E_{\beta}}}\Re\left(\frac{\partial\langle\phi(\bm{\theta})|}{\partial\bm{\theta}_n}\mathcal{H}|\phi(\bm{\theta})\rangle\right),
\end{split}
\label{Eq:C_vector}
\end{eqnarray}
where $E_{\beta}=1-2\delta\beta\langle\phi(\bm{\theta})|\mathcal{H}|\phi(\bm{\theta})\rangle$. The first term of $C_m$ can be computed by the quantum circuit in Fig.(2), and the second term can be calculated via the \emph{Median of Means} estimator and the Alg. \ref{Algorithm5}.

\textbf{Definition:} \emph{Median of Means estimator $\textbf{MM}_{N,K}(\cdot)$: Assume that the sample size $N=K[N/K]$, where $K$ is the number of subsamples and $[N/K]$ is the size of each subsample. We first randomly split the data into $K$ subsample and compute the mean using each subsample, which leads to estimators $X_1,X_2,...,X_K$ and each estimator is based on $[N/K]$ observations. The Median of Means estimator is defined as the median of all these estimator, i.e.,
\begin{align}
\textbf{MM}_{N,K}(X_k)=\textbf{Median}\{X_1,...,X_K\}.
\end{align}
}
Using the above estimator, one can efficiently estimate $C_m$ based on the \emph{Algorithm5}.

\begin{algorithm}
\caption{Estimating $\Re\left(\frac{\partial\langle\phi(\bm{\theta})|}{\partial\bm{\theta}_m}\mathcal{H}|\phi(\bm{\theta})\rangle\right)$}\label{algorithm}
\textbf{Input:} quantum states $|\phi(\bm{\theta})\rangle$ and $\frac{\partial|\phi(\bm{\theta})\rangle}{\partial\bm{\theta}_m}$, Hamiltonian $\mathcal{H}=\sum_{i=1}^{L}h_i$ and one ancillary qubit initialized to $|0\rangle$. \\
\textbf{Output:} Estimation value of $\Re\left(\frac{\partial\langle\phi(\bm{\theta})|}{\partial\bm{\theta}_m}\mathcal{H}|\phi(\bm{\theta})\rangle\right)$\\

(1) Initialize the quantum state

$$|\Psi_m\rangle=\frac{1}{\sqrt{2}}\left(|0\rangle||\phi(\bm{\theta})\rangle\rangle+|1\rangle\frac{\partial|\phi(\bm{\theta})\rangle}{\partial\bm{\theta}_m}\right),$$

and generate its Clifford samples via $U\in \mathbf{Cl}(2^n)$: $$S(\rho(\Psi_m);N)=\{\widehat{\rho}_1(\Psi_m),...,\widehat{\rho}_N(\Psi_m)\}.$$\\
(2) Split the $N$-samples into $K$ equally-sized parts and construct $K$ estimators

$$\widehat{\rho}_{(k)}=\frac{1}{[N/K]}\sum\limits_{l=(k-1)[N/K]+1}^{k[N/K]}\widehat{\rho}_l(\Psi_m)$$

\textbf{For}: $i=1$ to $L$ \textbf{do}\\
$$\widehat{o}_i(N,K)=\mathbf{MM}_{N,K}\left({\rm{Tr}}\left((\sigma^x\otimes h_i)\widehat{\rho}_{(k)}\right)\right)$$
\textbf{return}
$\Re\left(\frac{\partial\langle\phi(\bm{\theta})|}{\partial\bm{\theta}_m}\mathcal{H}|\phi(\bm{\theta})\rangle\right)\approx\sum\limits_{i=1}^{L}\widehat{o}_i(N,K)$
\label{Algorithm5}
\end{algorithm}

\subsection{Error analysis}
Now we provide the error analysis for using $|\phi(\bm{\theta})\rangle$ to approximate the Gibbs state $|\mu_{\beta}\rangle$. Taking the parameters $\delta\bm{\theta}=A^{-1}(\bm{\theta})C(\bm{\theta})$ into the loss-function $\mathcal{L}^2(\delta\beta)$, we obtain
\begin{align}
\mathcal{L}^2(\delta\beta)=\frac{1}{2}A^{-1}(\bm{\theta})C(\bm{\theta})C^{\dagger}(\bm{\theta})-\mathbf{d}\langle\mu_{\beta+\delta\beta}|\mathbf{d}|\mu_{\beta+\delta\beta}\rangle.
\end{align}

For the first term in $\mathcal{L}^2(\delta\beta)$, since $A(\bm{\theta})$ is a Hermitian matrix, it thus can be rewritten as $A(\bm{\theta})=\sum_{d=1}^D\lambda_d^A|\psi_d^A\rangle\langle\psi_d^A|$, where $\lambda_d^A$ denotes the $d$-th eigenvalue of $A(\bm{\theta})$ and $|\psi_d^A\rangle$ denotes the corresponding eigenvector. Then the vector $C(\bm{\theta})$ can be projected onto the basis $\{|\psi_d^A\rangle\}$, that is $C(\bm{\theta})=\sum_{d}\widetilde{C}_d|\psi_d^A\rangle$, where $\widetilde{C}_d=\sqrt{D}\sum_jC_j\langle j|\psi_d^A\rangle$. Therefore, the first term in $\mathcal{L}^2(\delta\beta)$ can be further calculated as $\sum_{m,s=1}^D\frac{1}{2\lambda_m^A}|\widetilde{C}_m\widetilde{C}_s^{\dagger}|$ which can be bounded by the theorem \ref{Theorem5}.

\begin{theorem}
Given the Hamiltonian $\mathcal{H}$ and the trial state $|\phi(\bm{\theta})\rangle=\prod_{d=1}^D\widetilde{U}_d(\theta_d)|\mu_0\rangle$, suppose each element $C_m$ in vector $C(\bm{\theta})$ is calculated via Eq.(\ref{Eq:C_vector}), then the norm of $C_m$ is bounded by
\begin{align}
\|C_m\|\leq\mathcal{O}\left(\frac{(\delta\beta)\lambda_{\max}}{\sqrt{E_{\beta}}}\right)
\end{align}
for $m=1,2,...,D$, where $\lambda_{\max}$ denotes the highest energy of $\mathcal{H}$.
\label{Theorem5}
\end{theorem}

\emph{Proof.} Since the relationship
\begin{align}
\frac{1}{\sqrt{E_{\beta}}}-1=\frac{(2\delta\beta)\langle\phi(\bm{\theta})|\mathcal{H}|\phi(\bm{\theta})\rangle}{\sqrt{E_{\beta}}(1+\sqrt{E_{\beta}})}
\end{align}
holds, the first term in Eq.(\ref{Eq:C_vector}) can be bounded by $\mathcal{O}(\frac{2(\delta\beta)\lambda_{\max}}{\sqrt{E_{\beta}}})$, where $\lambda_{\max}$ denotes the highest energy of $\mathcal{H}$. Furthermore, suppose $\mathcal{H}=\sum_{\lambda}|\psi_{\lambda}\rangle\langle\psi_{\lambda}|$, $|\phi(\bm{\theta})\rangle=\sum_{\lambda}a_{\lambda}|\psi_{\lambda}\rangle$ and $\frac{\partial|\phi(\bm{\theta})\rangle}{\partial\bm{\theta}_m}=\sum_{\lambda}b_{\lambda}|\psi_{\lambda}\rangle$, where $a_{\lambda},b_{\lambda}$ are complex values, the value $\Re\left(\frac{\partial\langle\phi(\bm{\theta})|}{\partial\bm{\theta}_n}\mathcal{H}|\phi(\bm{\theta})\rangle\right)$ can be evaluated as
\begin{align}
\Re\left(\frac{\partial\langle\phi(\bm{\theta})|}{\partial\bm{\theta}_n}\mathcal{H}|\phi(\bm{\theta})\rangle\right)=\sum\limits_{\lambda}\mathcal{R}(a_{\lambda}b_{\lambda}^{\dagger})\lambda.
\end{align}
Considering the complex-value coefficients $a_{\lambda},b_{\lambda}$ satisfy
\begin{align}
-\sum\limits_{\lambda}|a_{\lambda}b_{\lambda}^{\dagger}|\leq\sum\limits_{\lambda}\mathcal{R}(a_{\lambda}b_{\lambda}^{\dagger})\leq\sum\limits_{\lambda}|a_{\lambda}b_{\lambda}^{\dagger}|\leq1,
\end{align}
and $\sum_{\lambda}|a_{\lambda}b_{\lambda}^{\dagger}|\lambda\leq\lambda_{\max}$, we obtain $|C_m|\leq\mathcal{O}\left(\frac{(\delta\beta)\lambda_{\max}}{\sqrt{E_{\beta}}}\right).\Box$

According to the above theorem, we have  $|\widetilde{C}_m\widetilde{C}_s^{\dagger}|\leq \left(\sqrt{D}\max\{C_j\}\right)^2\leq\frac{D(\delta\beta)^2\lambda^2_{\max}}{E_{\beta}}$. Then the first term of $\mathcal{L}^2(\delta\beta)$ can be bounded by $\mathcal{O}\left(\frac{D(\delta\beta)^2\lambda^2_{\max}}{E_{\beta}}\sum_d\frac{1}{\lambda_m^A}\right)$.

For the second term $\mathbf{d}\langle\mu_{\beta+\delta\beta}|\mathbf{d}|\mu_{\beta+\delta\beta}\rangle$ in $\mathcal{L}^2(\delta\beta)$, it can be approximated by $(\delta\beta)^2\langle\mu_{\beta+\delta\beta}|\mathcal{H}|\mu_{\beta+\delta\beta}\rangle$ when $\delta\beta$ is small enough. Since $|\mu_{\beta+\delta\beta}\rangle$ represents the thermal state of $\mathcal{H}$, the state $|\mu_{\beta+\delta\beta}\rangle$ is close to the maximal entanglement state when the inverse temperature $\beta$ is close to $0$. In this situation,  $\langle\mu_{\beta+\delta\beta}|\mathcal{H}|\mu_{\beta+\delta\beta}\rangle$ euqals to $\sum_{i}\lambda_i/2^n$, where $\lambda_i$ is the $i$-th eigenvalue of $\mathcal{H}$. On the contrary, when the inverse temperature $\beta$ is large enough, $\langle\mu_{\beta+\delta\beta}|\mathcal{H}|\mu_{\beta+\delta\beta}\rangle$ is close to $\lambda_{\min}$. Then $\mathbf{d}\langle\mu_{\beta+\delta\beta}|\mathbf{d}|\mu_{\beta+\delta\beta}\rangle$ can be bounded by the interval $[(\delta\beta)^2\lambda_{\min},(\delta\beta)^2\sum_{i}\lambda_i/2^n]$.

Combining the above two estimations, we obtain that
$$\mathcal{L}^2(\delta\beta)\leq(\delta\beta)^2\left\|\frac{D\lambda^2_{\max}}{E_{\beta}}\sum_d\frac{1}{\lambda_m^A}-\lambda_{\min}\right\|,$$
and the proposed PVGS algorithm can provide a $\mathcal{O}(\beta\epsilon_3)$ approximation when the selected parameterized unitary satisfies
\begin{align}
\left\|\frac{D\lambda^2_{\max}}{E_{\beta}}\sum_d\frac{1}{\lambda_m^A}-\lambda_{\min}\right\|\leq\mathcal{O}(\epsilon_3^2),
\label{Eq:Measure}
\end{align}
where $\epsilon_3$ is a small positive value.

\section{Mean-Value-Clliford-Sampling (MVCS) algorithm
\label{Section6}
}
In this section, we focus on the last step in the \emph{PFCS-Algorithm}, that is, estimating the expectation values $\textmd{E}[V_i]$ and $\textmd{E}[W_i]$ for $i\in\{1,2,...,l\}$. Once again, for $i\in\{0,...,l-1\}$, random variables $V_i$ and $W_i$ are defined as $V_i=\exp(-d_{i,i+1}\mathcal{H})$, and $W_i=\exp(d_{i,i+1}\mathcal{H})$, where $d_{i,i+1}=(\beta_{i+1}-\beta_i)/2$. Therefore,
\begin{eqnarray}
\begin{split}
\textmd{E}[V_i]&=\sum\limits_{\mathbf{x}\sim|\mu_{\beta_i}\rangle}\frac{\exp(-\beta_i\mathcal{H}(\mathbf{x}))}{\mathcal{Z}(\beta_i)}\exp(-d_{i,i+1}\mathcal{H}(\mathbf{x}))\\
&=\langle\mu_{\beta_i}|\exp\left(-d_{i,i+1}\mathcal{H}\right)|\mu_{\beta_i}\rangle,
\end{split}
\end{eqnarray}
and
\begin{eqnarray}
\begin{split}
\textmd{E}[W_i]&=\sum\limits_{\mathbf{x}\sim|\mu_{\beta_{i+1}}\rangle}\frac{\exp(-\beta_i\mathcal{H}(\mathbf{x}))}{\mathcal{Z}(\beta_i)}\exp(d_{i,i+1}\mathcal{H}(\mathbf{x}))\\
&=\langle\mu_{\beta_{i+1}}|\exp\left(d_{i,i+1}\mathcal{H}\right)|\mu_{\beta_{i+1}}\rangle.
\end{split}
\end{eqnarray}
Naturally, to obtain approximations of mean values $\textmd{E}[V_i]$ and $\textmd{E}[W_i]$ with $\epsilon_4$ additive error for $i=1,2,...,l$, one need to invoke $\mathcal{O}(l/\epsilon_4)$ copies of state $|\mu_{\beta_i}\rangle$ and reflection $R=\left(2|\mu_{\beta_i}\rangle\langle\mu_{\beta_i}|-I\right)$ via using amplitude estimation algorithm \cite{Ashley2015Gibbs, Arunachalam2020Gibbs}. One of the disadvantages of this procedure is that the system error will be accumulated with the increasing of the estimated accuracy. In order to solve this problem and minimize the sampling complexity, we take the best advantage of the sampled Clifford samplings of $|\mu_{\beta_i}\rangle$ (generated in the first step) and utilize them to directly calculate the value of $\textmd{E}[V_i]$ and $\textmd{E}[W_i]$.

\subsection{Approximate $\exp\left(-d\mathcal{H})\right)$ by Chebyshev series}
Before proposing the elaborate steps, we introduce two theorems to approximate the operator $\exp\left(-d\mathcal{H}\right)$.

\begin{theorem}
Let $\delta_4,\epsilon_4\in(0,1)$ and real value function $f(\cdot)$ s.t. $\|f(x)-\sum_{k=0}^{K_{f}}a_kx^k\|<\epsilon_4/4$ for all $x\in[-1+\delta_4,1-\delta_4]$. Then there exists $\overrightarrow{c}\in\mathcal{R}^{2M_f+1}$ such that
\begin{align}
\left\|f(x)-\sum\limits_{m=-M_f}^{M_f}c_m\cos\left(\frac{m\pi x}{2}\right)\right\|\leq\epsilon_4
\end{align}
for all $x\in[-1+\delta_4,1-\delta_4]$, where $M_f=\max\left(2[\log\left(\frac{4\|a\|_1}{\epsilon_4}\right)\frac{1}{\delta_4},0]\right)$ and $\|\overrightarrow{c}\|_1\leq\|a\|_1$. Moreover $\overrightarrow{c}$ can be efficiently calculated on a classical computer in time $\textmd{poly}(K_f,M_f,\log(1/\epsilon_4))$.
\label{Theorem6}
\end{theorem}

Since the operator $\exp(-d\mathcal{H})$ is induced by the exponential function $f(x)=e^{-dx}$ that can be approximated by the truncated Taylor series:
\begin{align}
\left\|e^{-dx}-\sum_{k=0}^{K_f}\frac{(-dx)^k}{k!}\right\|\leq\frac{\epsilon_4}{4},
\end{align}
in which $K_f=\mathcal{O}\left(\frac{\log(d/\epsilon_4)}{\log\log(d/\epsilon_4)}\right)$,  according to theorem \ref{Theorem6}, one can efficiently calculate parameters $\overrightarrow{c}(d)\in\mathcal{R}^{2M_f+1}$ and obtain a Fourier approximation of $e^{-dx}$. To construct a bridge between Fourier approximation and Chebyshev series,
we define for $t\in\mathcal{R}^{+}$ and $\epsilon_4\in(0,1)$ the number $r(t,\epsilon_4)\geq t$ as the solution to the equation $\epsilon_4=\left(\frac{t}{r}\right)^r$, where $r\in(t,\infty)$. Literautre \cite{YuanSu2018QuantumSingular} indicated that for all $t>1$ one obtains
\begin{align}
r(t,\epsilon_4)=\Theta\left(t+\frac{\log(1/\epsilon_4)}{\log\log(1/\epsilon_4)}\right).
\end{align}
Using this estimation, we have the following theorem.

\begin{theorem}[A. Gilyen et al. \cite{YuanSu2018QuantumSingular}]
Let $t\in\mathcal{R}\backslash\{0\}$, $\epsilon_4\in(0,1/e)$, and let $R_t=[0.5r\left(\frac{e|t|}{2},\frac{5\epsilon_4}{4}\right)]$, then the following $2R_t$ degree polynomial satisfies
\begin{align}
\left\|\cos(tx)-J_0(t)+2\sum\limits_{k=1}^{R_t}(-1)^kJ_{2k}(t)T_{2k}(x)\right\|\leq\epsilon_4,
\end{align}
where $J_m(t)$ denotes the first kind Bessel function and $T_{2k}(x)$ denotes the first kind Chebyshev function.
\label{Theorem7}
\end{theorem}

Based on theorem \ref{Theorem7}, the function $e^{-dx}$ can be expanded by a
\begin{align}
R_{M_f}=\mathcal{O}\left(\frac{1}{\delta_4}\log\left(\frac{e}{\epsilon_4}\right)+\log\left(\frac{1}{\epsilon_4}\right)\right)
\end{align}
degree polynomial function, and the operator $\exp\left(-d\mathcal{H}\right)$ can thus be approximated by the operator
\begin{align}
2\sum\limits_{m=-M_f}^{M_f}\sum\limits_{k=0}^{R_m}c_m(d)(-1)^kJ_{2k}\left(\frac{m\pi}{2}\right)T_{2k}(\mathcal{H})
\end{align}
when all the eigenvalues of $\mathcal{H}$ belong to the interval $[-1+\delta_4,1-\delta_4]$. The index $R_m=[0.5r\left(\frac{e|m|}{2},\frac{5\epsilon_4}{4}\right)]$, $m$ takes value from the interval $[-M_f,M_f]$, and $M_f=\max\left(2[\log\left(\frac{4\|\overrightarrow{c}(d)\|_1}{\epsilon_4}\right)\frac{1}{\delta_4},0]\right)$

\subsection{Technical details of MVCS}
Once again, the estimation of $\textmd{E}[V_i]$ and $\textmd{E}[W_i]$ depend on efficiently extracting meaningful samples from the Gibbs state $|\mu_{\beta_i}\rangle$ and utilize these samples to
reflect the average property on the observable $\exp(\pm d_{i,i+1}\mathcal{H})$. From the above subsection, we know that the operator $\exp(-d_{i,i+1}\mathcal{H})$ can be approximated by the linear combinations of $\{\mathcal{H},\mathcal{H}^2,...,\mathcal{H}^{2R_{M_f}}\}$, therefore we can separately calculate the mean values $\langle\beta_i|\mathcal{H}^s|\beta_i\rangle$ for $s\in\{1,2,...,2R_{M_f}\}$ and combine them based on the corresponding coefficients.

For a fixed $\beta_i$ in the \emph{cooling schedule}, we have generated a $M$-scale \emph{Clifford Samples Set} of the state $|\mu_{\beta_i}\rangle$: $$S(|\mu_{\beta_i}\rangle,M)=\{\widehat{\rho}_1(\mu_{\beta_i}),...,\widehat{\rho}_{M}(\mu_{\beta_i})\}$$ in the first step of the \emph{PFCS-Algorithm}, where the sampling complexity $M$ is provided by Eq.(\ref{Eq:Sampling_complexity}), and these samples can be used to calculate $\textmd{E}[V_i]$ (or $\textmd{E}[W_i]$). To do this, we split $S(|\mu_{\beta_i}\rangle,M)$ into $K$ equally-sized parts and construct estimators $$\widehat{o}_s=\mathbf{MM}_{M,K}\left({\rm{Tr}}\left(\mathcal{H}^s\widehat{\rho}_{(k)}\right)\right)$$
for $s=1,2,...,2R_{M_f}$ and $k=1,2,...,K$. Finally, one can estimate $\textmd{E}[V_i]$ (or $\textmd{E}[W_i]$) via combining each estimators $\widehat{o}_s$ with the corresponding coefficients.

\section{Complexity Analysis}
\subsection{Computational Complexity}
The overall structure of the proposed PFCS algorithm is:\\
(1) Use the CSBS algorithm to compute a decent \emph{cooling schedule} $(\beta_1,...,\beta_l)$ of length $l$.\\
(2) Use the PVGS algorithm to generate the Gibbs states $|\mu_{\beta_i}\rangle$.\\
(3) Use the MVCS algorithm to estimate the expectations $\textmd{E}[W_i]$ and $\textmd{E}[V_i]$, then multiply these estimates to obtain an estimation of $\textmd{E}[W]$ and $\textmd{E}[V]$, and output their ratio as the final estimate.

Now we analyze the time complexity in each step. For the CSBS algorithm (step 1), according to the Theorem 3.4 in literature \cite{Arunachalam2020Gibbs}, the length of cooling schedule $l=\sqrt{q\ln n}$ suffices to estimate
$$S[W_i]=S[V_i]=\frac{\mathcal{Z}(\beta_i)\mathcal{Z}(\beta_{i+1})}{\mathcal{Z}(\frac{\beta_i+\beta_{i+1}}{2})}\leq15$$
for every $i\in[l]$, where $q=\ln(\mathcal{Z}(\beta)/\mathcal{Z}(\beta_0))$. In the CSBS algorithm, we perform binary search with precision $\alpha=1/2n$ over the domain that is contained in $[0,\beta]$, which implies that the number of steps for determining an inverse temperature $\beta_i$ is at most $\log(2n\beta)$. Then the total number of binary searches in all steps is $l\log(2n\beta)=\sqrt{q\ln n}\log(2n\beta)$. Each step in binary research invokes the Alg. \ref{Algorithm2} to estimate the estimation variance with additive error $\epsilon_2$ and failure probability $\delta_2$. Combining theorem \ref{Theorem2}, the Clifford sampling complexity of step (1) is
\begin{align}
\mathcal{O}\left(\frac{\sqrt{q\ln n}\log(2n\beta)\log(1/\delta_2)}{\epsilon^2_2}\right),
\end{align}
where $\epsilon_2$ is the additive error for estimating variances $\textmd{S}[V_i], \textmd{S}[W_i]$ and $\delta_2$ is the failure probability.

For the PVGS algorithm (step 2), we utilize a $D$-depth quantum circuit to approximate the Gibbs state $|\mu_{\beta}\rangle$ at inverse temperature $\beta$. The fundamental complexity comes from constructing the $D\times D$ matrix $A(\bm{\theta})$ as well as the $D\times 1$ vector $C(\bm{\theta})$. According to the Eq.(\ref{Eq:Hardamard-Test}), one can efficiently estimate the element $\widehat{A}_{n,m}$ via $\mathcal{O}(1/\epsilon_3^2)$ quantum samplings. The value of $C_m$ is estimated by Clifford samplings extracted from the state
$$
|\Psi_m\rangle=\frac{1}{\sqrt{2}}\left(|0\rangle||\phi(\bm{\theta})\rangle\rangle+|1\rangle\frac{\partial|\phi(\bm{\theta})\rangle}{\partial\bm{\theta}_m}\right)
$$
by using
Alg. \ref{Algorithm5}. To estimate the sampling complexity, let $X$ be a random variable with variance $\sigma^2$. Then $K$ independent sample means of size $[N/K]=\mathcal{O}(1/\epsilon_3^2)$ suffice to construct a median of means estimator $\mu_i(N,K)$ that obeys
\begin{align}
\mathbf{Pr}\left(\left|\mu_i(N,K)-\textmd{E}[X]\right|\geq\epsilon_3\right)\leq2e^{-K/2}
\end{align}
for all $\epsilon_3>0$. If we assign $X={\rm{Tr}}\left((\sigma^x\otimes h_i)\widehat{\rho}_{(k)}\right)$ ($k=1,...,K$), the parameters $K$ and $N$ are selected such that this general statement ensures
\begin{align}
\mathbf{Pr}\left(\left|\widehat{o}_i(N,K)-\textmd{E}[X]\right|\geq\epsilon_3\right)\leq2e^{-K/2}=\delta_3/L,
\end{align}
in which $\delta_3$ indicates the failure probability and $L$ represents the number of terms in the Hamiltonian $\mathcal{H}$. Therefore, the parameter $K=\mathcal{O}(\log(L/\delta_3))$ and the total sampling complexity for estimating each $C_m$ is $\mathcal{O}\left(\frac{\log(L/\delta_3)}{\epsilon_3^2}\right)$.
Finally, it takes
\begin{align}
\mathcal{O}\left(\frac{n\beta(D^2+D\log(L/\delta_3))}{\epsilon_3^2}\right)
\end{align}
Clifford samplings to recover the Gibbs state $|\mu_{\beta}\rangle$ of a general physical Hamiltonian $\mathcal{H}$.
\begin{figure*}[htb]
\centering
  \centering
  \subfigure[]{
  \includegraphics[width=0.47\textwidth]{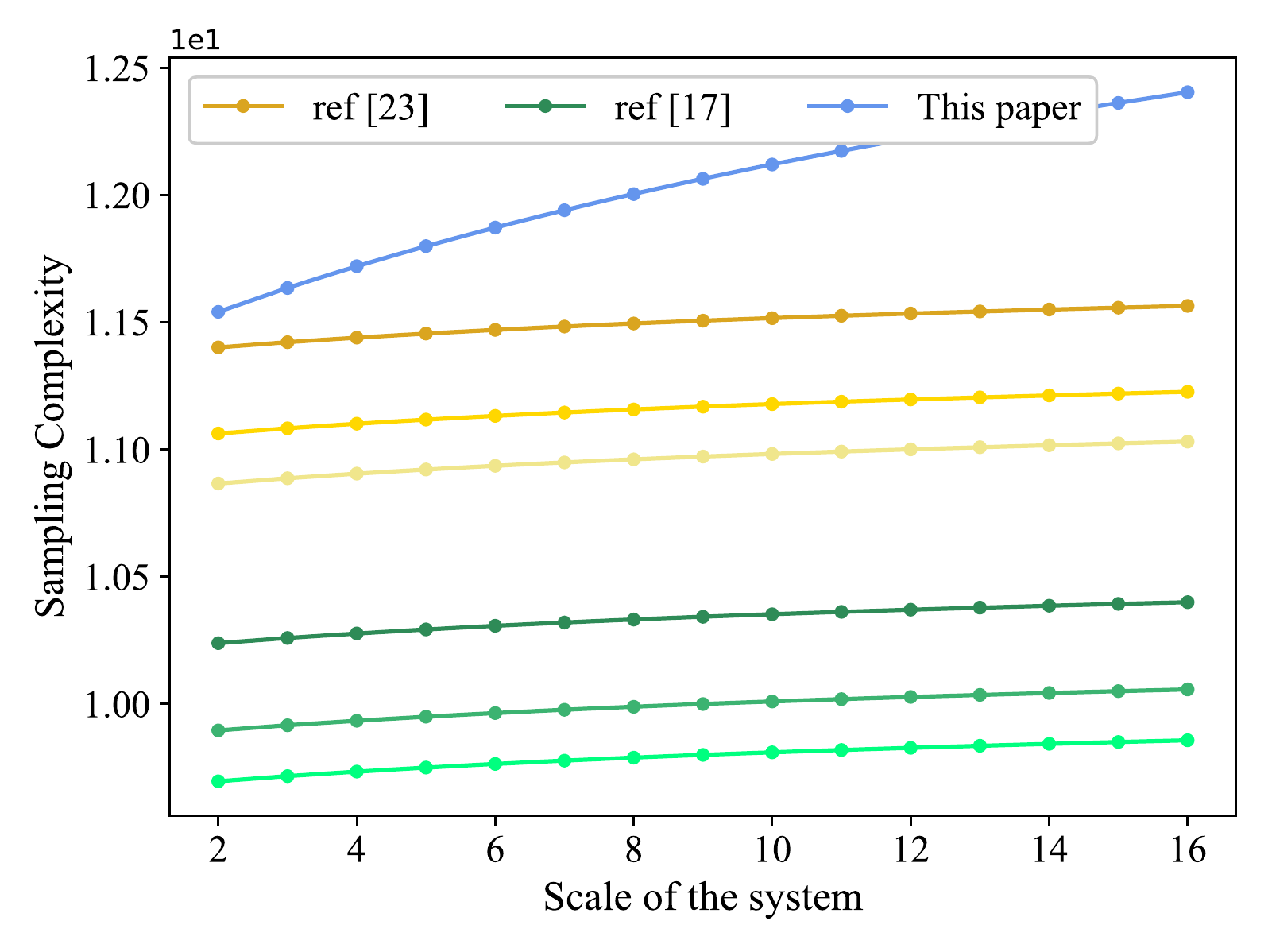}
  }
  \centering
  \subfigure[]{
  \includegraphics[width=0.47\textwidth]{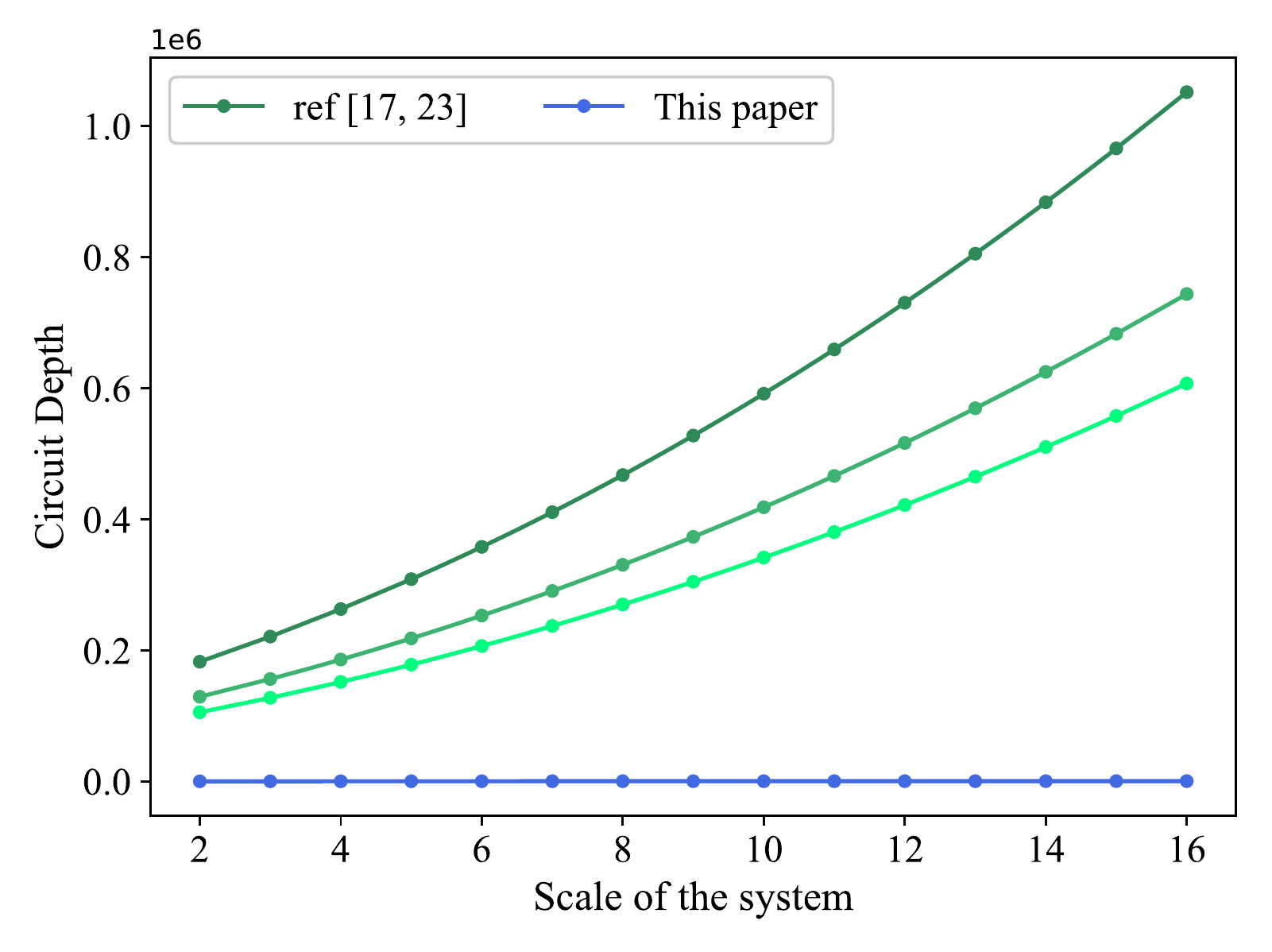}
  }
  \centering
  \subfigure[]{
  \includegraphics[width=0.47\textwidth]{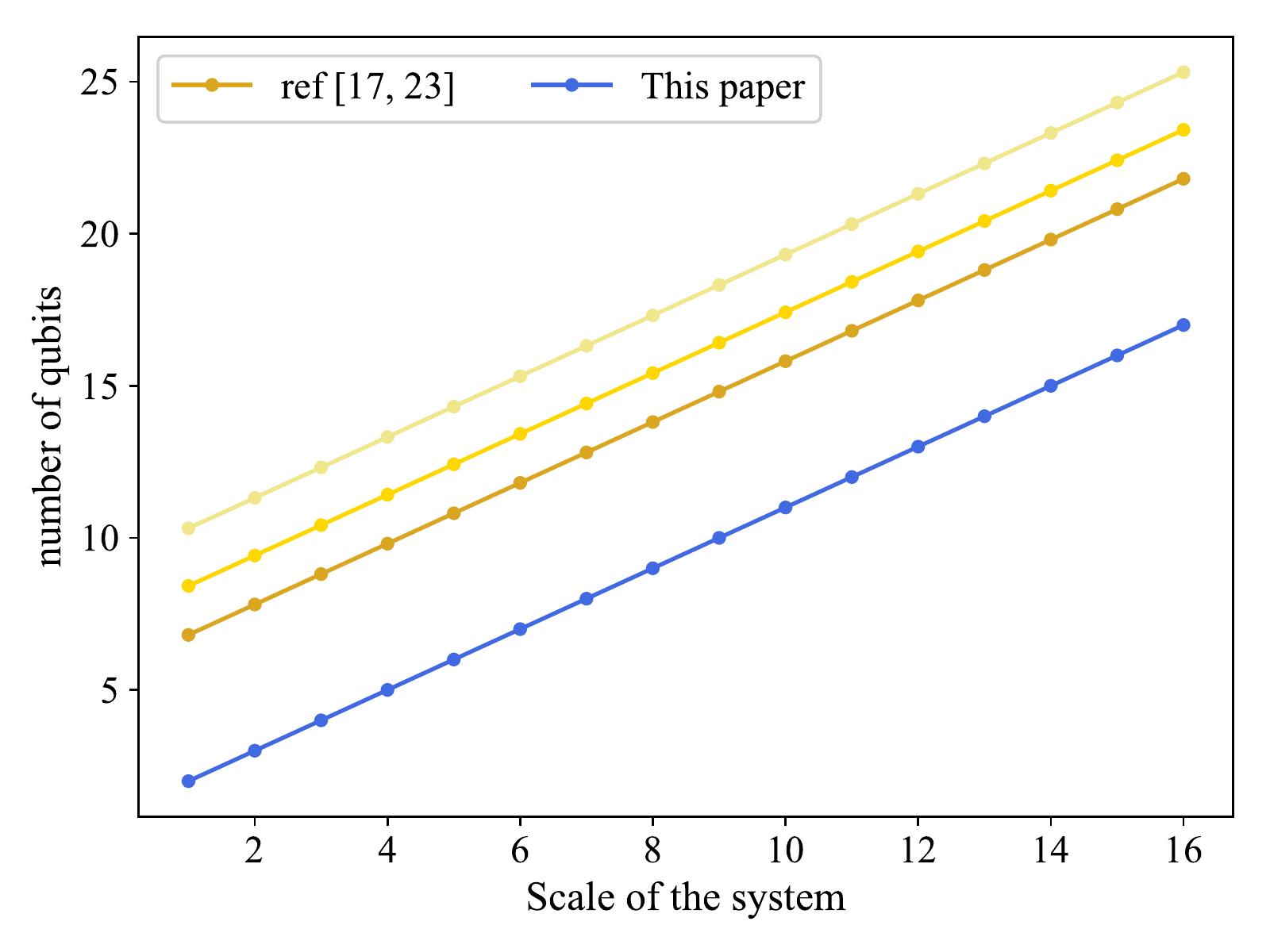}
  }
  \DeclareGraphicsExtensions.
  \caption{Computational resource comparison in terms of (a) `Sampling Complexity', (b) `Circuit Depth' and (c) `Number of qubits'. Here, we choose $\beta=2$, spectral gap $\Delta\in\{10^{-2}, 10^{-3}, 10^{-4}\}$ and $\epsilon=10^{-2}$, in which lighter curves correspond to smaller $\Delta$. Besides that, we set parameters $|W|=n^2$, $D=10$ and $L=10$ in estimating the computational resource of `Sampling complexity'.}
  \label{Fig:GroundEnergy}
\end{figure*}

\begin{figure}[htb]
  \begin{center}
  \includegraphics[width=0.5\textwidth]{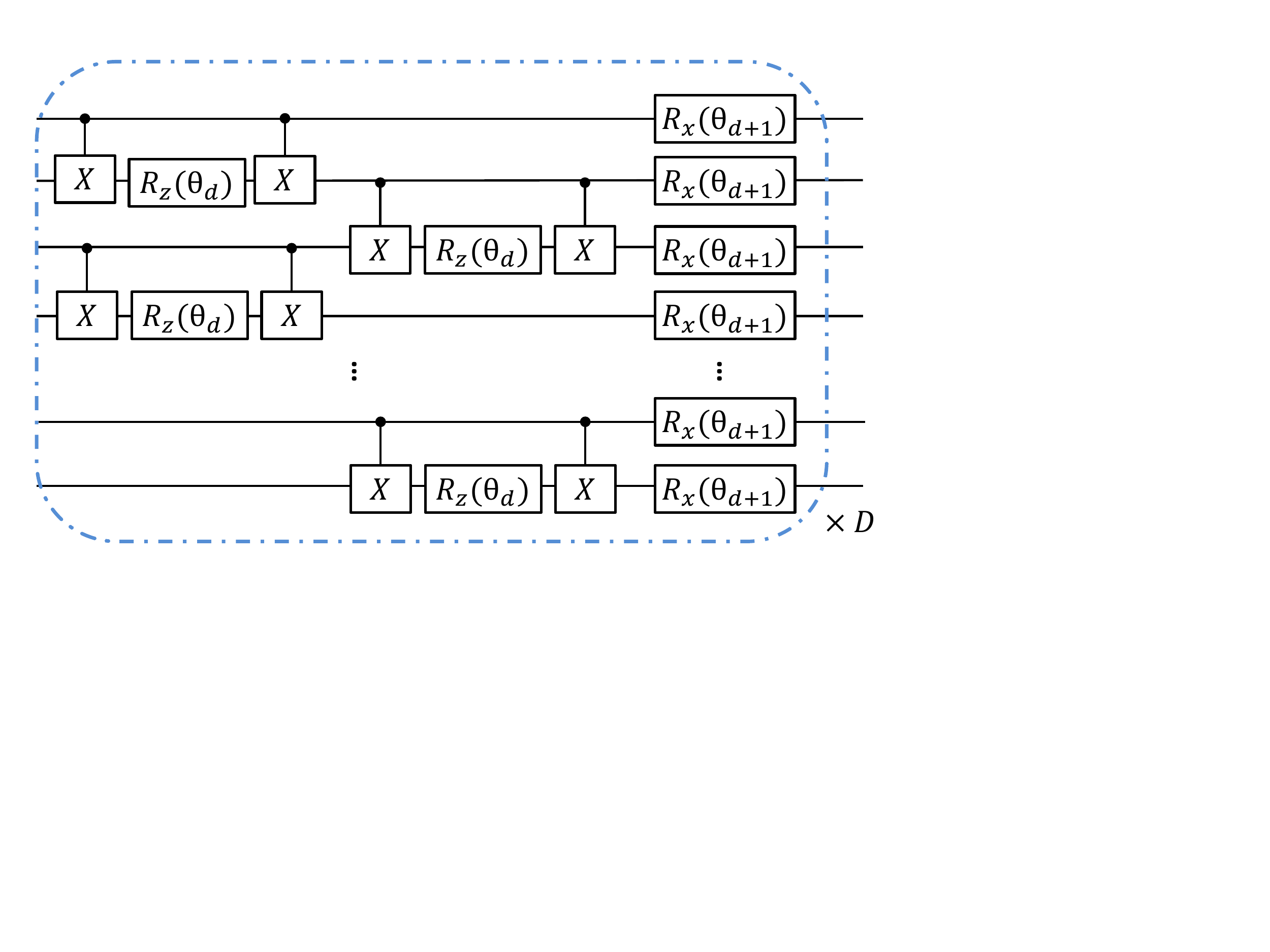}\\
  \caption{The quantum circuit for simulating Gibbs states of a diagonal Hamiltonian. Here, the block depth $D=5$ and the number of qubits $n=10$.}
  \end{center}
  \label{Fig:circuit1}
\end{figure}

\begin{figure*}[htb]
\centering
  \centering
  \subfigure[]{
  \includegraphics[width=0.47\textwidth]{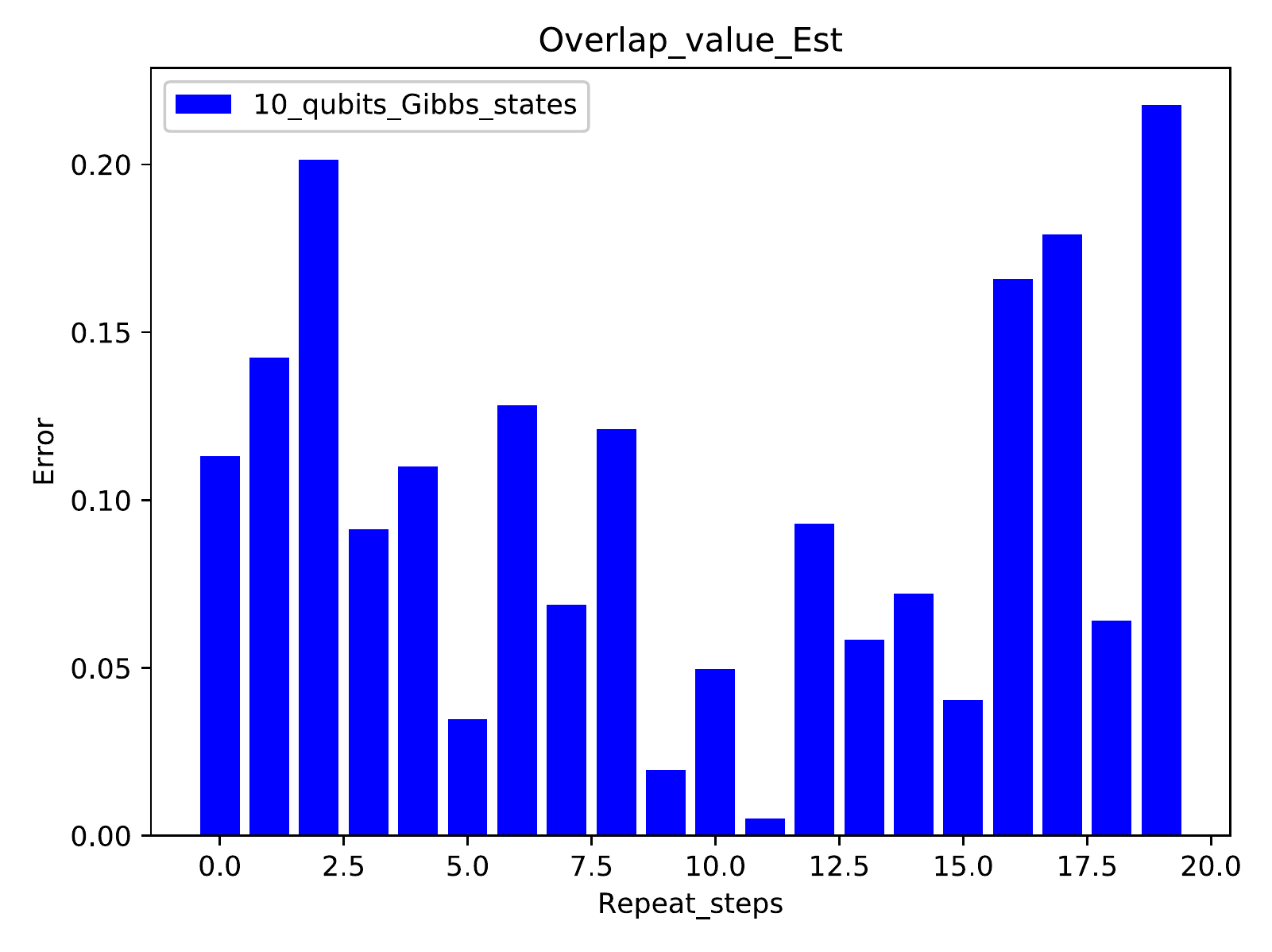}
  }
  \subfigure[]{
  \includegraphics[width=0.47\textwidth]{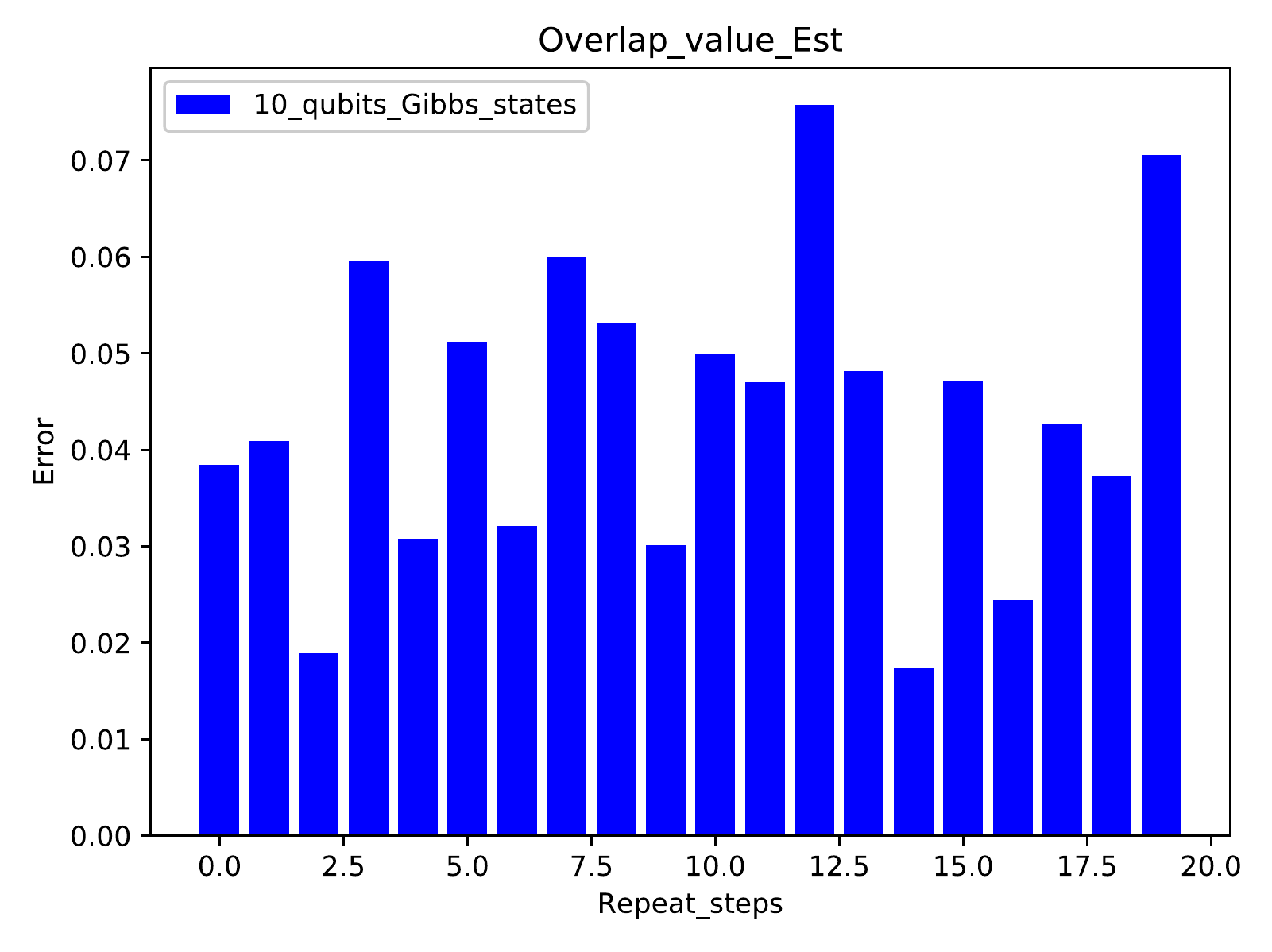}
  }
  \DeclareGraphicsExtensions.
  \caption{The error of overlap estimation between $10$-qubit quantum states $|\mu_{\beta_i}\rangle$ and $|\mu_{\beta_j}\rangle$ by using the PVGS and Alg. \ref{Algorithm2}. (a) The Clifford sampling complexity $M_s=100$, and the average additive error of 20 experiments is 0.10. (b) The Clifford sampling complexity $M_s=1000$, and the average additive error of 20 experiments is $0.03$. }
  \label{Fig:GibbsState}
\end{figure*}

\begin{figure*}[htb]
\centering
  \centering
  \subfigure[]{
  \includegraphics[width=0.47\textwidth]{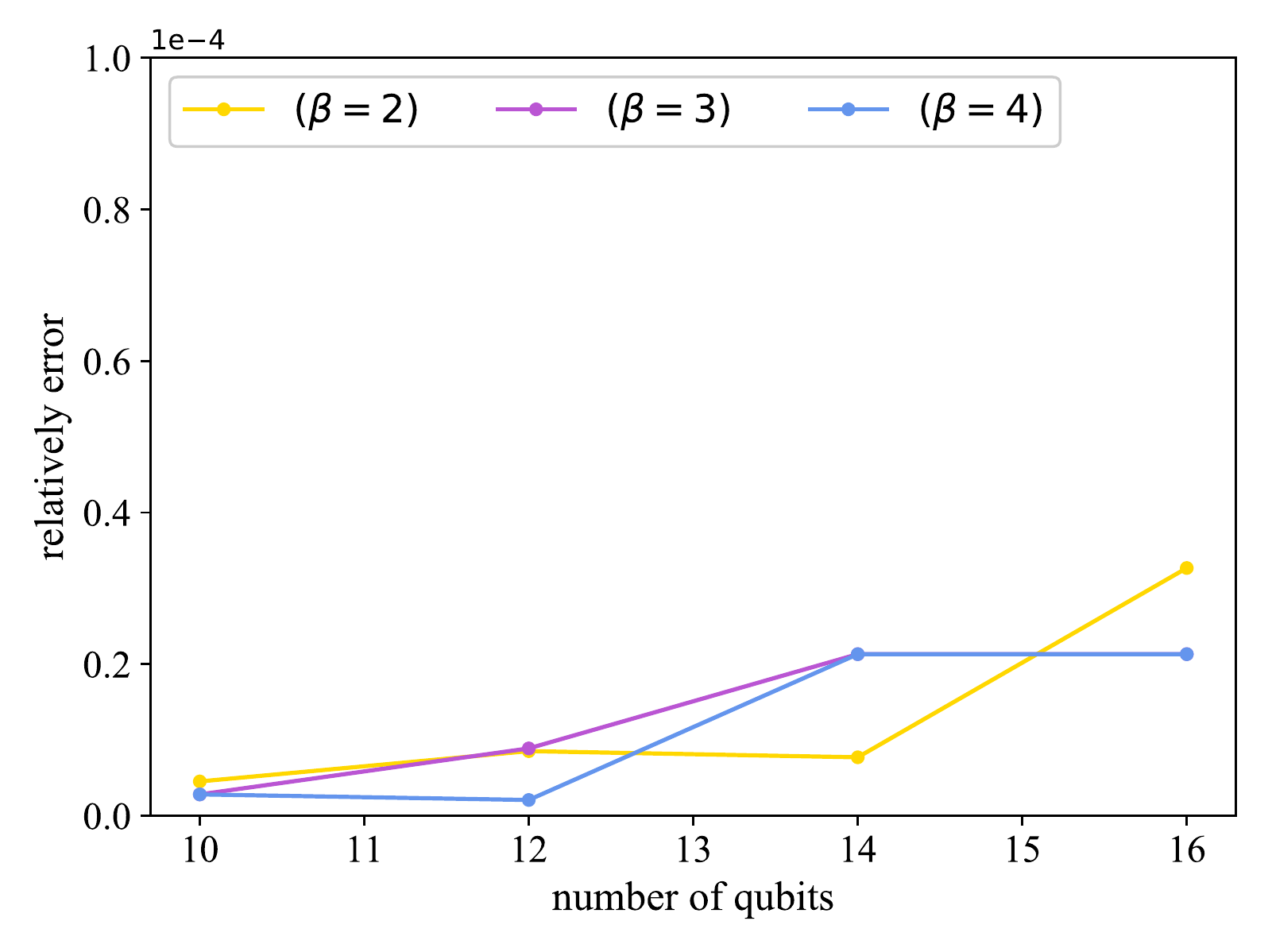}
  }
  \centering
  \subfigure[]{
  \includegraphics[width=0.47\textwidth]{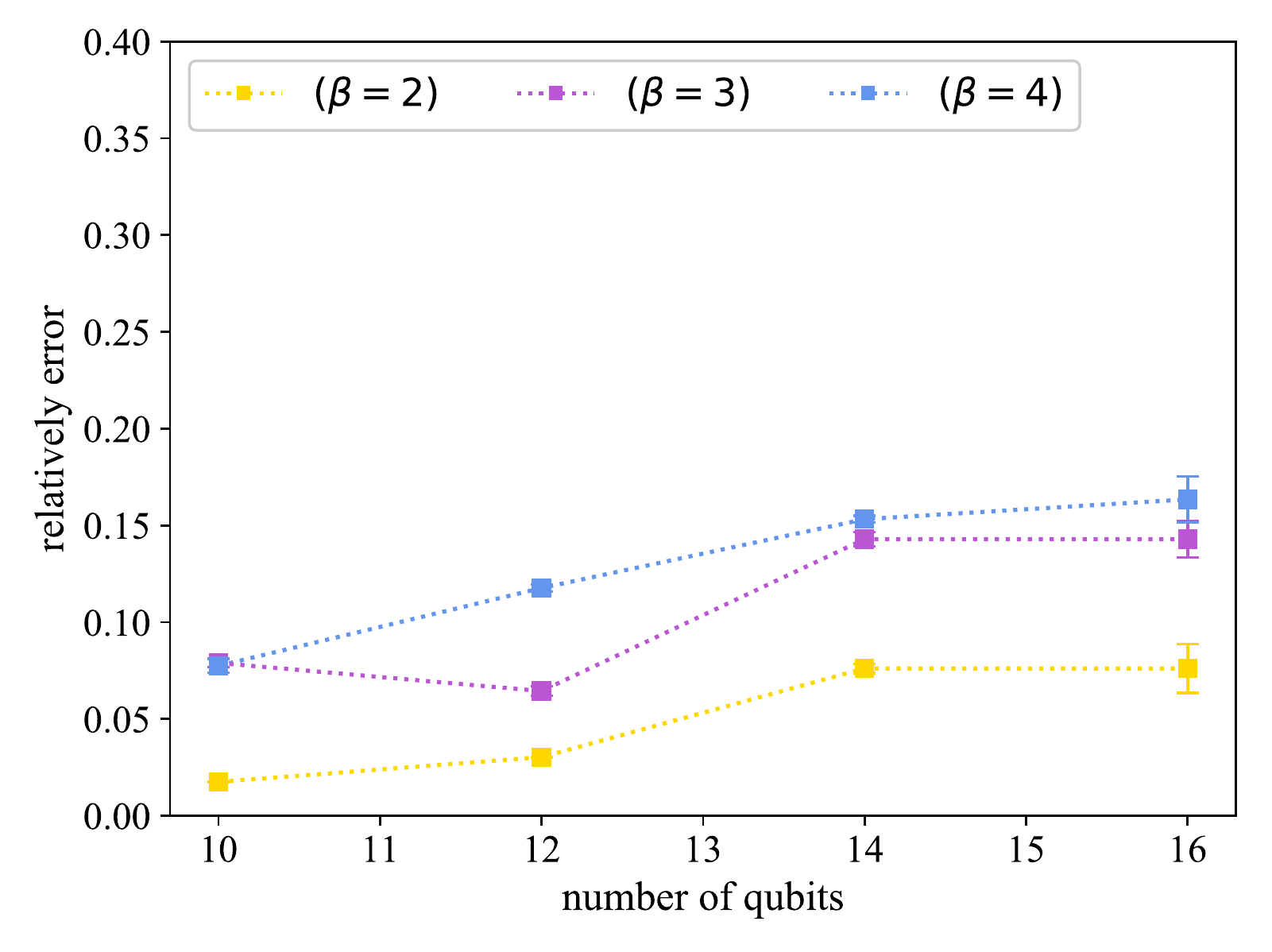}
  }
  \DeclareGraphicsExtensions.
  \caption{The relative error by using the \emph{PFCS-Algorithm} to calculate partition function $\mathcal{Z}(\beta)$ of a diagonal Hamiltonian for $\beta=2,3,4$. The left graph (a) indicates the relative error by using exact value of $\textmd{E}[V_i]$ and $\textmd{E}[W_i]$ via implementing infinite Clifford samplings, and the right graph (b) indicates the relative error by using $M_s=1000$ Clifford samplings.}
  \label{Fig:GroundEnergy}
\end{figure*}

\begin{figure*}[ht]
\centering
  \centering
  \subfigure[]{
  \includegraphics[width=0.47\textwidth]{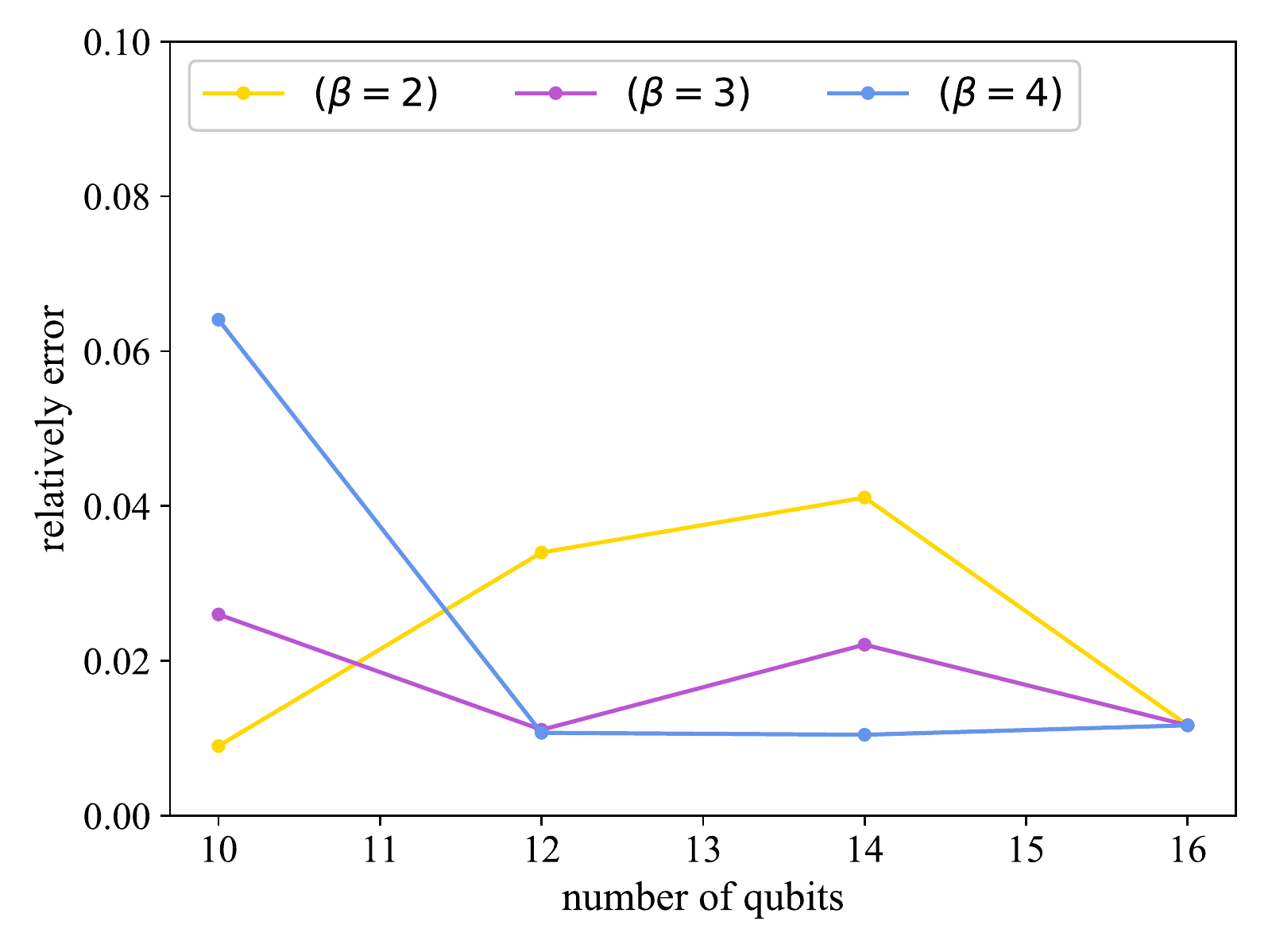}
  }
  \centering
  \subfigure[]{
  \includegraphics[width=0.47\textwidth]{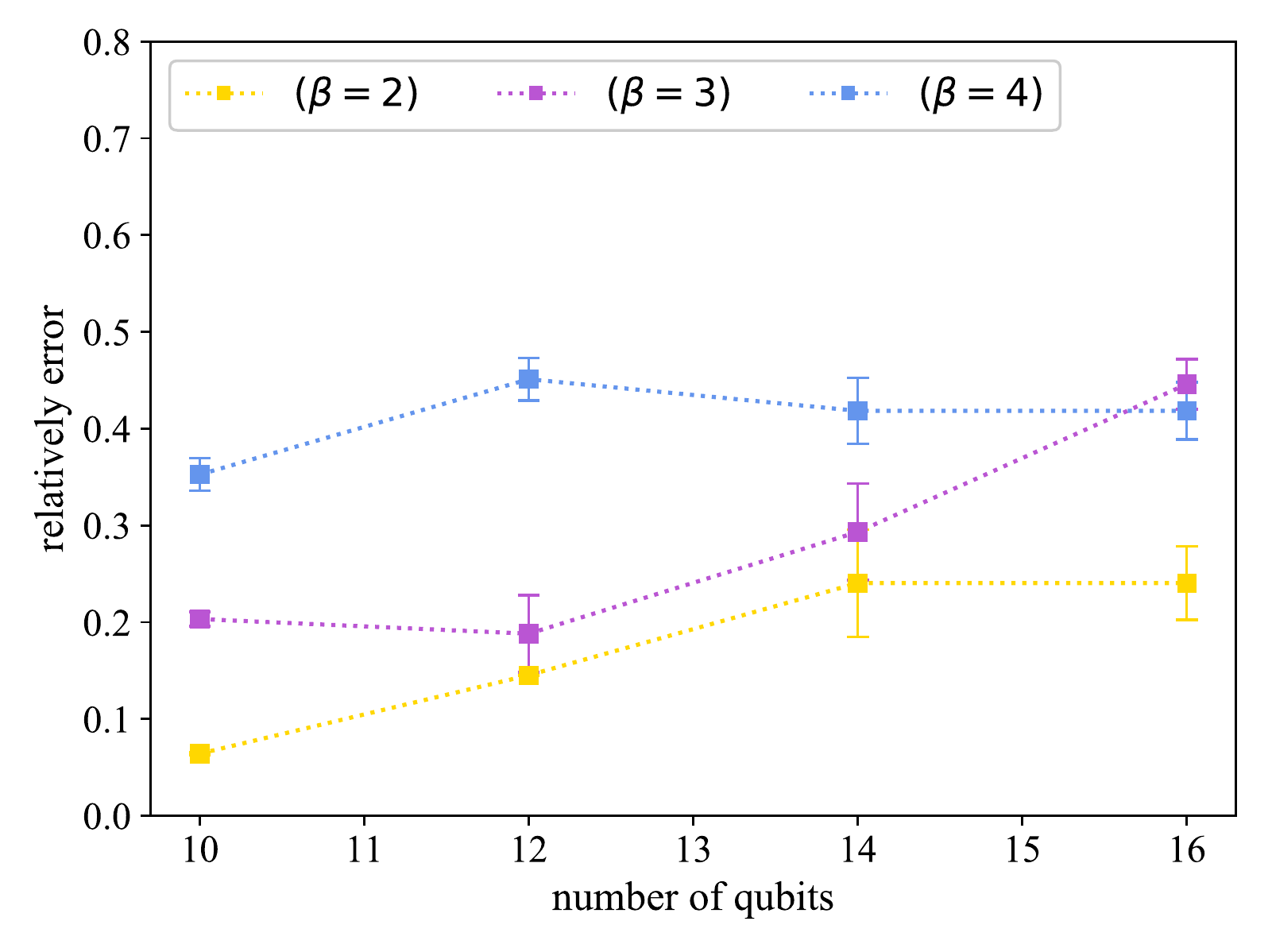}
  }
  \DeclareGraphicsExtensions.
  \caption{The relative error by using the \emph{PFCS-Algorithm} to calculate partition function $\mathcal{Z}(\beta)$ of Ising model with transverse field for $\beta=2,3,4$. The left graph (a) indicates the relative error by using exact value of $\textmd{E}[V_i]$ and $\textmd{E}[W_i]$ via implementing infinite Clifford samplings, and the right graph (b) indicates the relative error by using $M_s=1000$ Clifford samplings.}
  \label{Fig:GroundEnergy}
\end{figure*}

\begin{figure*}[ht]
\centering
  \centering
  \subfigure[]{
  \includegraphics[width=0.47\textwidth]{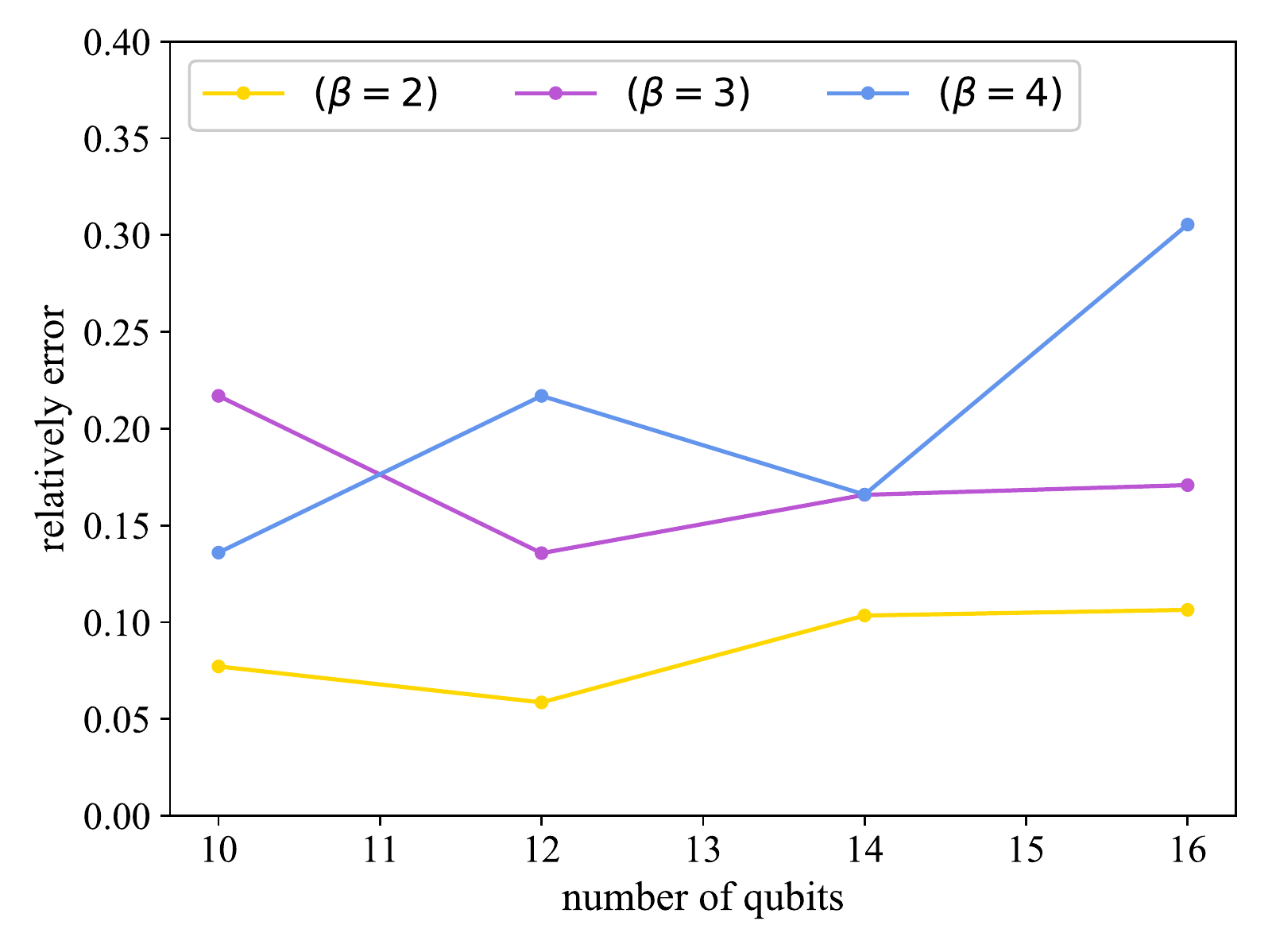}
  }
  \centering
  \subfigure[]{
  \includegraphics[width=0.47\textwidth]{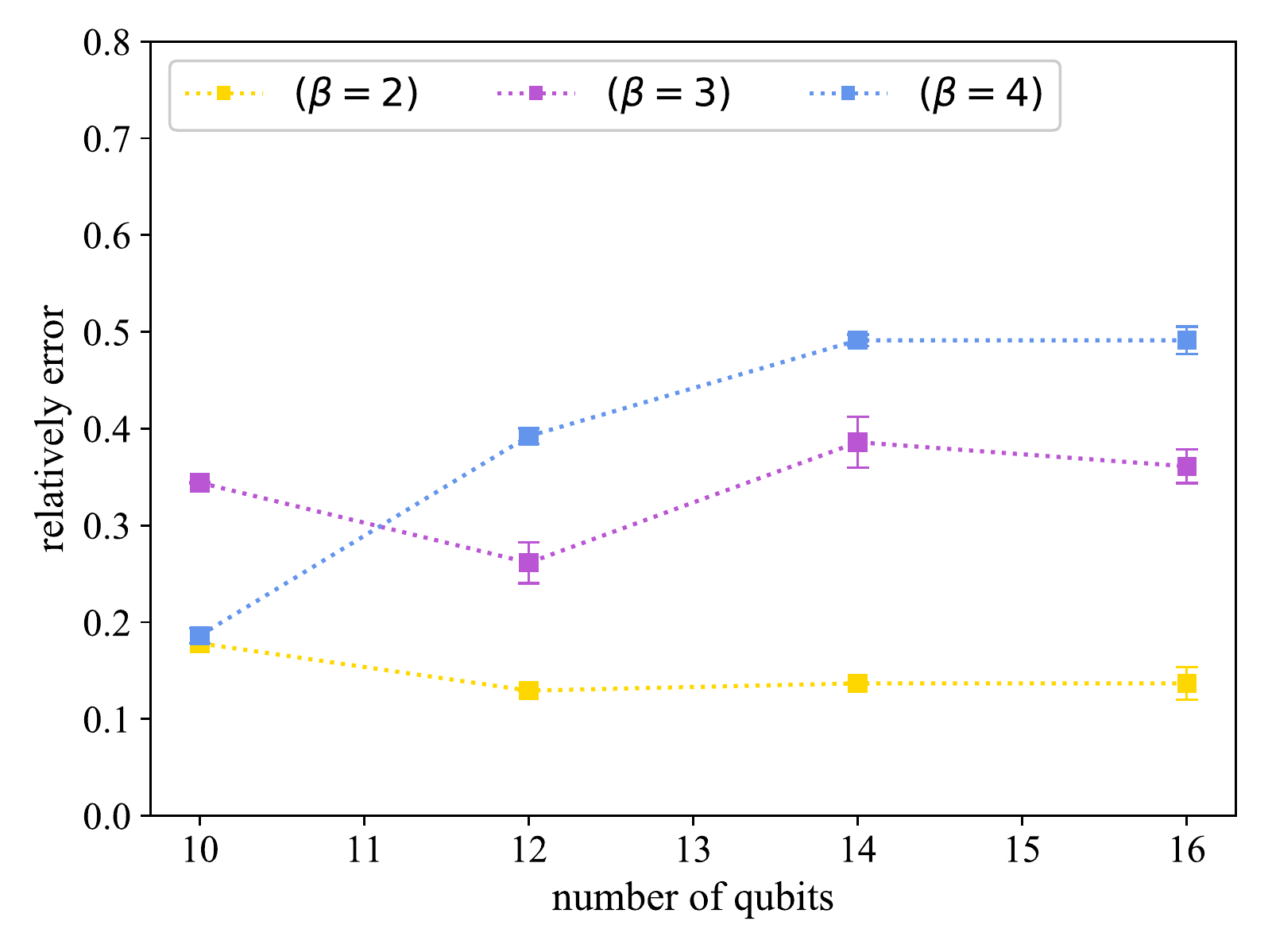}
  }
  \DeclareGraphicsExtensions.
  \caption{The relative error by using the  \emph{PFCS-Algorithm} to calculate partition function $\mathcal{Z}(\beta)$ of 2D-Hubbard model for $\beta=2,3,4$. The left graph (a) indicates the relative error by using exact value of $\textmd{E}[V_i]$ and $\textmd{E}[W_i]$ via implementing infinite Clifford samplings, and the right graph (b) indicates the relative error by using $M_s=1000$ Clifford samplings.}
  \label{Fig:GroundEnergy}
\end{figure*}
\begin{table*}
\begin{tabular}{ |p{3.5cm}|p{3cm}|p{4cm}|p{4cm}|}
 \hline
 \multicolumn{4}{|c|}{Sampling complexity and Resource requirements} \\
 \midrule[1pt]
 S. Arunachalam \cite{Arunachalam2020Gibbs}:($\mathbf{d}$)& \quad Cooling schedule      &\qquad\ Gibbs Sampling   & \quad Mean-Value estimation \\
 \hline
 $\bullet$ Sampling complexity  & \quad\quad$\mathcal{O}\left(l\log(1/\epsilon)\right)$
                            &\quad\quad\quad $\mathcal{O}\left(Bl/\sqrt{\Delta}\right)$
                            &\quad\quad\quad$\mathcal{O}\left(\sqrt{B}l\log(1/\epsilon)\right)$\\
 \hline
 $\bullet$ Qubits & \quad\quad $n+\log(1/\epsilon)$    & \quad\quad\quad $n+\log(1/\Delta)$   &\quad\quad\quad $n+\log(1/\epsilon)$ \\
 \hline
 $\bullet$ Circuit depth & \quad\quad $\mathcal{O}(|W|/\epsilon\sqrt{\Delta})$    &  \quad\quad \quad$\mathcal{O}(|W|/\sqrt{\Delta})$   & \quad\quad\quad $\mathcal{O}(|W|/\epsilon\sqrt{\Delta})$\\
 \midrule[1pt]

 A. Montanaro \cite{Ashley2015Gibbs}:($\mathbf{d}$)& & &\\
 \hline
 $\bullet$ Sampling complexity  & \quad$\mathcal{O}\left(Bl^2\log(1/\epsilon)\right)$
                            &\quad\quad\quad $\mathcal{O}\left(Bl/\sqrt{\Delta}\right)$
                            &\quad\quad\quad$\mathcal{O}\left(Bl\log(1/\epsilon)\right)$\\
 \hline
 $\bullet$ Qubits & \quad\quad $n+\log(1/\epsilon)$    & \quad\quad\quad $n+\log(1/\Delta)$   &\quad\quad\quad $n+\log(1/\epsilon)$ \\
 \hline
 $\bullet$ Circuit depth & \quad\quad $\mathcal{O}(|W|/\epsilon\sqrt{\Delta})$    &  \quad\quad \quad$\mathcal{O}(|W|/\sqrt{\Delta})$   & \quad\quad\quad $\mathcal{O}(|W|/\epsilon\sqrt{\Delta})$\\
 \midrule[1pt]
 \hline
     This paper:($\mathbf{d}$ and $\mathbf{g}$) & & &\\
 \hline
  $\bullet$ Sampling complexity $\uparrow$ & \quad\quad$\mathcal{O}\left(l/\epsilon^2\right)$
                            & $\mathcal{O}\left(n\beta(D^2+D\log(L/\delta))/\epsilon^2\right) $
                            &$\mathcal{O}\left(BR_{M_f}\log(L)\log(1/\delta)/\epsilon^2\right)$\\
 \hline
 $\bullet$ Qubits $\downarrow$& \quad\quad\quad\quad $n$   & \quad\quad\quad\quad $(n+1)$  & \quad\quad\quad\quad\quad $n$ \\
 \hline
 $\bullet$ Circuit depth $\downarrow$ & \quad\quad $D+C(k)$    &  \quad\quad\quad\quad $D+C(k)$   & \quad\quad\quad\quad $D+C(k)$\\
 \midrule[1pt]
\end{tabular}
\label{Table1}
\caption{The comparation between the proposed algorithm and previous works \cite{Ashley2015Gibbs,Arunachalam2020Gibbs} in terms of Sampling complexity and Resource requirements. Here `$\mathbf{d}$' indicates a diagonal Hamiltonian and  `$\mathbf{g}$' indicates a general Hamiltonian with off-diagonal elements. The parameter $B$ is the upper bound of $\textmd{Var}(X_i)/\textmd{E}^2[X_i]$, where $X_i\in\{V_i,W_i\}$, $\Delta$ is the spectral gap of the Markov chain, and $|W|$ is the circuit depth for quantum walk operators. According to the literature \cite{SzegedyDepth}, the parameter $|W|=\textmd{poly}(n)$ on sparse graphs. The function $C(k)$ is the average depth of a $k$-qubit Clifford gate, in detail, $C(1)=1$ and $C(k)=k^2/\log(k)$ for $1<k\leq n$. The signal $\downarrow$ marks the reduced quantum resource in this paper, and vice versa.}
\end{table*}

For the MVCS algorithm (step 3), there are approximately $\mathcal{O}(L^{R_{M_f}})$ Pauli terms in the operators $\exp(\pm d\mathcal{H})$ under the assumption that $\mathcal{H}=\sum_{s=1}^Lh_s$. Noting that the Clifford sampling method provides an estimation of $\textmd{E}[V_i]$ (and $\textmd{E}[V_i]$) with an additive error $\widetilde{\epsilon}_4$, that is, $|\widehat{V}_i-\textmd{E}[V_i]|\leq\widetilde{\epsilon}_4$. To obtain a relative estimation, the additive error should be adjusted to $\epsilon_4=|\textmd{E}[V_i]|\widetilde{\epsilon}_4$. Considering that the expectation of $V_i$ equals to $\frac{\mathcal{Z}(\frac{\beta_i+\beta_{i+1}}{2})}{\mathcal{Z}(\beta_i)}$ which is bounded by a constant value $1/\sqrt{c_2}$,
a scale of  
\begin{align}
M=\mathcal{O}\left(\frac{BR_{M_f}\log(L)\log(1/\delta_4)}{\epsilon^2_4}\right)
\label{Eq:Sampling_complexity}
\end{align}
Clifford samplings suffice to provide an $\epsilon_4=\mathcal{O}(\widetilde{\epsilon}_4)$-relative estimation, where $B$ denotes the upper bound of $\textmd{Var}(X_i)/\textmd{E}^2[X_i]$ and $X_i\in\{V_i,W_i\}$ (also see theorem \ref{Theorem1}). Then we obtain the estimations of $V_i$ and $W_i$ such that
\begin{align}
1-\epsilon_4/(2l)\leq\frac{V_i}{\textmd{E}[V_i]}\leq1+\epsilon_4/(2l),
\end{align}
and
\begin{align}
1-\epsilon_4/(2l)\leq\frac{W_i}{\textmd{E}[W_i]}\leq1+\epsilon_4/(2l)
\end{align}
with the probability of $1-1/(20l)$ ($\delta_4=1/(20l)$) as well as $\epsilon_4=\mathcal{O}(1/l)$. After that, we utilize ratios of the lower and upper bounds to characterize the ratio $W_i/V_i$ from below and above and employ the union bound to obtain
\begin{align}
(1-\epsilon_4/(2l))^{2l}\leq\frac{\prod_i(W_i/V_i)}{\prod_i(\textmd{E}[W_i]/\textmd{E}(V_i))}\leq(1+\epsilon_4/(2l))^{2l}.
\end{align}
Since the relationships $(1-2\epsilon_4)\leq(1-\epsilon_4/(2l))^{2l}$ and $(1+2\epsilon_4)\geq(1+\epsilon_4/(2l))^{2l}$ hold, we obtain
\begin{align}
(1-2\epsilon_4)\frac{\mathcal{Z}(\beta)}{\mathcal{Z}(\beta_0)}\leq\prod_i(W_i/V_i)\leq(1+2\epsilon_4)\frac{\mathcal{Z}(\beta)}{\mathcal{Z}(\beta_0)},
\end{align}
that is a $(2\epsilon_4)$-relative estimation of $\frac{\mathcal{Z}(\beta)}{\mathcal{Z}(\beta_0)}$.

Putting everything together, our algorithm needs
\begin{align}
\mathcal{O}\left(\frac{l\log(1/\delta)(\log(2n\beta)+BR_{M_f}\log(L))}{\epsilon^2}\right)
\end{align}
samples of Gibbs state, and all the Gibbs states require
\begin{align}
\mathcal{O}\left(\frac{n\beta(D^2+D\log(L/\delta))}{\epsilon^2}\right)
\end{align}
Clifford sampling complexity when we assume $\epsilon_2=\epsilon_3=\epsilon_4=\epsilon$ and $\delta_2=\delta_3=\delta_4=\delta$.
\subsection{Comparison with previous work}
Here, we provide the computational resources comparison between the proposed algorithm and previous arts, and the results are listed as \emph{Table1}. According to the upper bound of computational resources, we visualize the three kind of quantum resources by selecting the spectral gap $\Delta=\{10^{-2}, 10^{-3}, 10^{-4}\}$ and $\epsilon=10^{-2}$ in Fig.3, in which lighter curves correspond to smaller $\Delta$. In theses three subgraphs, blue curves indicate the required quantum resources by using the proposed algorithm, and yellow curves, green curves represent the quantum resources by literatures \cite{Arunachalam2020Gibbs,Ashley2015Gibbs}, respectively. From this visualization, we can clearly obtain the advantages and disadvantages of our scheme in these three resources.

From the comparison, we first indicate that previous schemes mainly concentrate on diagonal Hamiltonians which encode all the information on its diagonal, that is $\mathcal{H}(\mathbf{x})=\sum_{i=1}^nx_i+\sum_{i,j}^nx_ix_j$ whose eigenvector $\mathbf{x}=x_1x_2...x_n$ ($x_i\in\{0,1\}$). And our algorithm can be applied to both diagonal Hamiltonians and general Hamiltonians, in which the fundamental gap lies on the sampling efficiency between the Clifford sampling and the $\{0,1\}^{\otimes n}$ random sampling on estimating $\textmd{E}[V_i]$ and $\textmd{E}[W_i]$. 

To estimate the mean value of an algorithm $\mathcal{A}$, the \emph{Mean-Value-Estimation} algorithms \cite{Ashley2015Gibbs, MeanValue1, MeanValue2} generally introduced a unitary acting on $n+1$ qubit, that is
$$U|\mathbf{x}\rangle|0\rangle=|\mathbf{x}\rangle\left(\sqrt{1-\phi(\mathbf{x})}|0\rangle+\sqrt{\phi(\mathbf{x})}|1\rangle\right),$$
where $\mathbf{x}\in\{0,1\}^n$ and $\phi(\mathbf{x})$ is the output by algorithm $\mathcal{A}$ when measurement outcome $\mathbf{x}$ is received. Then applying the amplitude estimation several times, one obtain the mean value $\textmd{E}[v(\mathcal{A})]$, where $v(\mathcal{A})$ is the random variable corresponding to the value computed by $\mathcal{A}$. This procedure is extremely suitable for sampling from a Gibbs state of diagonal Hamiltonians, since the diagonal Hamiltonian  $\mathcal{H}(\mathbf{x})$ encodes all the eigenvalues on its diagonal elements. Here, this algorithm essentially samples from the Gibbs state $|\mu_{\beta}\rangle=\sum_{\mathbf{x}}\frac{e^{-\beta \mathcal{H}(\mathbf{x})/2}}{\sqrt{\mathcal{Z}(\beta)}}|\mathbf{x}\rangle|\mathbf{x}\rangle
$ via computational basis, and using these samples to recover the mean value, therefore the above algorithm does not efficiently work on general off-diagonal quantum Hamiltonians.

After that, the proposed scheme reduces the number of qubits used in the whole algorithm. In our algorithm, at most $(n+1)$-qubit suffice to complete the whole procedure and provide an estimation of the partition function. According to the \emph{Table1}, the amplitude estimation based algorithms require relatively large number of qubits when parameter $1/\Delta$ is extremely large (e.g. $\mathcal{O}(2^n)$), and our algorithm is more suitable for intermediate-scale quantum devices.

Finally, we successfully reduce the depth of required quantum circuits. Our algorithm requires a $(D+C(n))$-depth quantum circuit, and the selection of $D=\log(n), C(1)=\mathcal{O}(1)$ promise less noise is accumulated.

\section{Simulation Results}
\subsection{Simulation results for the Alg. \ref{Algorithm2} and the PVGS}
Here, we validate the correctness of Alg. \ref{Algorithm2} and the PVGS algorithm by analyzing the diagonal  Hamiltonian
$$\mathcal{H}(\mathbf{x})=\sum_{i=1}^nx_i+\sum_{i,j}^nx_ix_j,$$
where $\mathbf{x}=x_1x_2...x_n$ ($x_i\in\{0,1\}$). To do this, we first generate its Gibbs states $|\mu_{\beta_i}\rangle, |\mu_{\beta_j}\rangle$ at inverse temperatures $\beta_i,\beta_j\in[0,2]$ by using the PVGS algorithm, then we predict the quantum state overlap $|\langle\mu_{\beta_i}|\mu_{\beta_j}\rangle|^2$ with the help of Alg. \ref{Algorithm2}. In this subsection, we utilize a relatively small-scale scenario that $n=10$ to validate the correctness of these two algorithms, and corresponding results are illustrated as Fig.5.

In each group of experiment, we randomly select $20$ different inverse temperature pairs $(\beta_i,\beta_j)$, and we utilize the Hamiltonian Variational (HV) ansatz in the PVGS algorithm, that is
$$|\phi(\bm{\theta})\rangle=\prod\limits_{d=1}^D\exp\left(-i\theta_d\mathcal{H_A}\right)\exp\left(-i\theta_{d+1}\mathcal{H_B}\right)|\mu_0\rangle,$$
in which $\mathcal{H}_A=\sum_{i}^n\sigma_{i}^z\sigma_{i+1}^z$ and $\mathcal{H}_B=\sum_{i}^n\sigma_i^x$, and the corresponding quantum circuit is illustrated as Fig.4. We implement Clifford sampling for $M_s=100$ times (see Fig.5.a) and $M_s=1000$ times (see Fig.5.b) to test the relationship between the additive error $\epsilon$ and sampling times $M_s$. From these $20$ groups of experiments, we find that $M_s=\mathcal{O}(1/\epsilon^2)$ that obeys the upper bound proposed in theorem \ref{Theorem2}.

\subsection{Estimating Partition functions}
Then we utilize the proposed \emph{PFCS-Algorithm} to calculate the partition function of diagonal Hamiltonians, 1D-Ising model with transverse field and $2D$ Fermi-Hubbard model, and the simulation results are illustrated as Fig.6-8. Considering the performance of the algorithm is mainly affected by the sampling complexity, we separately provide the estimation results via using infinite Clifford samplings (see Fig.6-8.a) and using $M_s=1000$ Clifford samplings (see Fig.6-8.b).

For the transverse field $1$D-Ising model,
\begin{align}
\mathcal{H}_{I}=\sum\limits_{i=1}^n\sigma_i^z\sigma_{i+1}^z+\sum\limits_{i=1}^n\sigma_i^x.
\end{align}
We test scenarios from $n=10$ to $n=16$, and the inverse temperature $\beta$ takes value from $\{2,3,4\}$. In the CSBS algorithm, we utilize $M_s$ Clifford samplings in each iteration, and the PVGS algorithm selects HV ansatz to approximate the Gibbs state at each inverse temperature $\beta_i$ for $i\in[l]$. Finally, in the MVCS algorithm, $\textmd{E}[V_i]$ and $\textmd{E}[W_i]$ are calculated by $M_s$ Clifford samplings. The simulation results for $M_s=\infty$ and $M_s=1000$ 
are illustrated as Fig.7, in which the $y$-axis represents the relative error $\epsilon_t$ between the theory value $\mathcal{Z}(\beta)$ and the result provided by the \emph{PFCS-Algorithm}, that is, $$(1-\epsilon_t)\mathcal{Z}(\beta)\leq\widehat{\mathcal{Z}}(\beta)\leq(1+\epsilon_t)\mathcal{Z}(\beta),$$
where $\widehat{\mathcal{Z}}(\beta)$ is the output of \emph{PFCS-Algorithm}. According to the simulation results, we find that the relative error $\epsilon_t$ increases to approximately $0.4$ with the increasing of the inverse temperature and the number of qubits.

Finally, we utilize \emph{PFCS-Algorithm} to compute partition function of physical systems with long Pauli strings, whose length increases with the grid size. We simulate the iconic 2D Fermi-Hubbard model with at most $8$ sites (16 qubits), and the target Hamiltonian is defined as
\begin{align}
\mathcal{H}=-t\sum\limits_{\langle i,j\rangle, \sigma}\left(a^{\dagger}_{i\sigma}a_{j\sigma}+a^{\dagger}_{j\sigma}a_{i\sigma}\right)+U\sum\limits_in_{i\uparrow}n_{i\downarrow},
\end{align}
where $a_{i\sigma}^{\dagger}$ and $a_{i\sigma}$ are fermionic creation and annihilation operators, $n_{i\uparrow}=a^{\dagger}_{i\uparrow}a_{i\uparrow}$ and similarly for $n_{i\downarrow}$. The notation $\langle i,j\rangle$ in the first sum associates sites that are adjacent in a $n_a\times n_b$ grid, and $\sigma\in\{\uparrow, \downarrow\}$. We utilize the Jordan-Wigner transformation to map each fermionic mode to a qubit. In detail, the hopping term between qubits $i$ and $j$ ($i<j$) maps to one qubit operator via
\begin{align}
a^{\dagger}_{i\sigma}a_{j\sigma}+a^{\dagger}_{j\sigma}a_{i\sigma}\mapsto\frac{1}{2}\left(X_iX_j+Y_iY_j\right)Z_{i+1}...Z_{j-1},
\end{align}
and the on-site term maps to a qubit operator via
\begin{align}
n_{i\uparrow}n_{i\downarrow}\mapsto\frac{1}{4}\left(I-Z_i\right)\left(I-Z_j\right).
\end{align}
Here, we utilize the \emph{PFCS-Algorithm} to approximate the partition function $\mathcal{Z}(\beta)$ of 2D Hubbard models. We test scenarios that $t=1$, $U=2$ and $n_a\times n_b=2\times k$, where $k\in[5,8]$. The simulation results are illustrated as Fig.8. Once again, the left three solid lines (yellow, purple and blue) indicate the relative error $\epsilon_t$ via $M_s=\infty$ Clifford samplings, and the right three dotted lines reflect the results by using $M_s=1000$ Clifford samplings. With the increasing of the inverse temperature and the number of qubits, the relative error $\epsilon_t$ of $16$-qubit $\mathcal{Z}(\beta)$ will increase to approximately $0.5$ at $\beta=4$.

\section{Conclusion}
A pressing open question for quantum computing in the Noisy Intermediate-Scale Quantum (NISQ) era is whether a shallow-depth quantum circuit can demonstrate quantum advantages in solving problems of practical significance. Recent outstanding works in this area include solving linear algebra~\cite{Sergey2018Science} and Boolean function~\cite{Sergy2021QuantumAdvantage} problems. In this paper, we established a quantum-classical hybrid algorithm for estimating the partition function of a general Hamiltonian, named as the \emph{PFCS-Algorithm}. 

To estimate the partition function, previous works require $\mathcal{O}(1/\epsilon\sqrt{\Delta})$-depth quantum circuits, where $\Delta$ is the minimum spectral gap of stochastic matrices and $\epsilon$ is the multiplicative error~\cite{Arunachalam2020Gibbs, Ashley2015Gibbs}. Through the use of novel Clifford sampling techniques, the \emph{PFCS-Algorithm} proposed in this paper only requires a $\mathcal{O}(1)$-depth quantum circuit with an $(n+1)$-qubit quantum device to provide a comparable $\epsilon$ approximation of an $n$-qubit partition function. Such a substantial reduction in the circuit complexity is achieved by an increase in the sampling complexity, which requires the $\mathcal{O}(1)$-depth quantum circuit to repeat  $\mathcal{O}(n/\epsilon^2)$ times to yield the same $\epsilon$ approximation. 
We then applied the \emph{PFCS-Algorithm} to a variety of Hamiltonians, including a classical Hamiltonian, the transverse-field Ising model, and the 2D Hubbard model, covering interesting application scenarios, such as molecular and Fermionic systems. In conclusion, the proposed \emph{PFCS-Algorithm} algorithm is not only significant in theory, but also delivers application values especially in the NISQ era.

\end{document}